\documentclass[showpacs,amsmath,amssymb]{revtex4}

\usepackage{graphicx}% Include figure files
\usepackage{dcolumn}% Align table columns on decimal point
\usepackage{bm}% bold math
\usepackage{braket}% braket
\usepackage{url}
\begin{document}

\preprint{APS/123-QED}

\title{Effect of Elastic Vibrations on Normal Head on Collisions of Isothermal Spheres}% Force line breaks with \\

\author{Ryo Murakami}%
 \email{ryo-mura@yukawa.kyoto-u.ac.jp}
% \altaffiliation[Also at ]{}%Lines break automatically or can be forced with \\
\author{Hisao Hayakawa}
\affiliation{Yukawa Institute for Theoretical Physics, Kyoto University, Sakyo-ku, Kyoto, Japan, 606-8502}%
%\author{Charlie Author}
% \homepage{http://www.Second.institution.edu/~Charlie.Author}
%\affiliation{
%Second institution and/or address\\
%This line break forced% with \\
%}%
\date{\today}% It is always \today, today,
             %  but any date may be explicitly specified

\begin{abstract}
We numerically investigate head on collisions of isothermal visco-elastic spheres. 
We find that the restitution coefficient oscillates against the impact speed if the solid viscosity inside the sphere is small enough.
We confirm that the oscillation arises from the resonance between the duration of contact and the eigen-frequencies of the sphere.
This oscillation disappears if there exists the strong solid viscosity in spheres.
We also find that a sinusoidal behavior of the restitution coefficient against the initial phase in the eigenmodes for collisions between a thermally activated sphere and a flat wall.
As a result, the restitution coefficient can exceed unity if the impact speed of the colliding sphere is nearly equal to or slower than the thermal speed.
We have confirmed the existence of the fluctuation theorem for impact processes through our simulation.
\end{abstract}

% PACS, the Physics and Astronomy
                             % Classification Scheme.
%\keywords{Suggested keywords}%Use showkeys class option if keyword
                              %display desired
\maketitle

\section{\label{sec:intro} INTRODUCTION}

%The collision of viscoelastic bodies can be seen in granular systems. 
A granular material is a collection of discrete solid grains characterized by the loss of energy during collisions between grains.
Granular materials such as sands and powders are commonly observed not only on the earth~\cite{herrmann1998, hinrichsen2004}
but also on the other planets and satellites such as Mars~\cite{bridges2012} and the moon~\cite{kumar2013} as well as  planetary disks~\cite{weidenschilling1980, nakagawa1981, tanaka2005, dullemond2005, brauer2008}.
%[
%Bridges12=N. T. Bridges, F. Ayoub, J-P. Avouac, S. Leprince, A. Lucas and S. Mattson, Nature 485,  339 (2012).
%YOU SHOULD LOOK FOR THE OTHER PAPERS ON THIS ISSUE.
%]
Granular material behaves as an unusual liquid or a solid or a gas depending on its setup~\cite{jaeger1996}. 
% [jeager96= H. M. Jeager, S. R. Nagel and R. P. Behringer, Rev. Mod. Phys. {\bf 68}, 1259 (1996). ]
It is, of course, important to control the behavior of granular materials in engineering and industry~\cite{castellanos2005, tomas2007_1, tomas2007_2, tykhoniuk2007, bertram2008}.
The rich behavior of granular materials mainly arises from inelastic collisions between grains which are the results of
competition among attractive, repulsive and dissipative interactions between contacting grains~\cite{hinrichsen2004}.
Therefore, to know the properties of collision processes of grains plays a key role to understand the physics of granular materials. 

Collisions between grains are characterized by the restitution coefficient $e$, the ratio of the rebound speed to the impact speed, which is usually assumed to be $0 \leq e \leq 1$.
Although the majority of textbooks of elementary mechanics states that $e$ can be treated as a material constant, 
recent studies on collision dynamics reveal that the restitution coefficient behaves more complicated:
The restitution coefficient depends on the impact speed~\cite{bridges1984,johnson1985,kuwabara1987,brilliantov1996,morgado1997},
%[ 
%Bridges84=F. G. Bridges, A. Hatzes, and D. N. C. Lin, Nature (London) 309, 333 (1984).
%Johnson85= K. L. Johnson, Contact Mechanics (Cambridge Univ. Press, Cambridge, 1985).
%Kuwabara87=G. Kuwabara and K. Kono, Jpn. J. Appl. Phys. 26, 1230 (1987).
%Brilliantov96=  N. V. Brilliantov, F. Spahn, J. M. Hertzsch, and T. Po\"{o}schel, Phys. Rev. E 53, 5382 (1996).
%Morgado97= W. A. Morgado and I. Oppenheim: Phys. Rev. E 55, 1940  (1997).
%]
the restitution coefficient can exceed unity in oblique collisions~\cite{louge2002,kuninaka2004,hayakawa2004,shen2012},
%[
%Louge02=M. Y. Louge and M. E. Adams, Phys. Rev. E 65, 021303(2002).
%Kuninaka04= H. Kuninaka and H. Hayakawa, Phys. Rev. Lett. 93, 154301 (2004).
%Hayakawa04= H. Hayakawa and H. Kuninaka, Phase Transit. 77, 889 (2004).
%Shen12= C Shen, Y Wang, S Wang, Y Liu, R Liu, A Vourlidas, B. Miao, P. Ye, J. Liu and Z. Zhou, Nature Phys. 8, 923 (2012).
%]
and the restitution coefficient can be negative in oblique collisions~\cite{saitoh2010,mueller2012}.
%[
%Saitoh10= K. Saitoh, A. Bodrova, H. Hayakawa, and N. V. Brilliantov, Phys. Rev. Lett. 105, 238001 (2010).
%Mueller12 = P. M\"{u}ller, D. Krengel, and T. P\"{o}schel, Phys. Rev. E 85, 041306 (2012). 
%]
Recently, M\"uller \textit{et al}. performed a remarkable experiment and observed step-wise behavior of the restitution coefficient against the impact speed~\cite{mueller2013}, where a steel sphere of the diameter 6 mm bounces repeatedly off the glass plate.
They suggested that match or mismatch between the vibration frequency of the glass plate and the free flight time is responsible for this stair-wise behavior.
%It should be noted that these anomalous rebounds can be observed in collisions between macroscopic bodies characterized only by the repulsive interaction between grains.

Collisions, however, between small grains such as fine powders, known as cohesive dry powders, are strongly affected by attractive surface force in particular for slow collisions~\cite{johnson1985,castellanos2005,awasthi2006,brilliantov2007,
suri2008,kuninaka2009,saitoh2010,kim2011,tanaka2012,kuninaka2012}.
%[
%Awasthi06=A. Awasthi, S. C. Hendy, P. Zoontjens, and S. A. Brown, Phys. Rev. Lett. 97, 186103 (2006).
%Brilliantov07=] N. V. Brilliantov, N. Albers, F. Spahn, and T. P\"{o}schel,  Phys. Rev. E 76, 051302 (2007).
%Suri08= M. Suri and T. Dumitrica, Phys. Rev. B 78, 081405(R) (2008).
%kim11= S. Kim, Phys. Rev. E 83, 041302 (2011).
%Tanaka12= H. Tanaka, K. Wada, T. Suyama and S. Okuzumi,  Prog. Theor. Phys. Suppl. No. 195, 101 (2012).
%]
Note that
this attractive force between fine powders is in-avoidable, because it originates in the inter-atomic forces such as van-der Waals force. 
As a result, a variety of processes in collisions of fine powders can be observed depending on their impact speeds~\cite{harbich2000}.
Nano-powders fragment into atoms~\cite{pettersson1993, chatelet1992} or several large components or bury themselves on walls~\cite{yamamura1994, kenny2002} for sufficiently high speed impacts.
On the other hand, colliding powders are coalesced if the impact speed is too slow as in adsorptions of powders on walls \cite{cheng1994, saitoh2009} and clustering in freely falling granular streams~\cite{waitukaitis2011}.
It is, however, possible to reduce the attractive force by the surface coating of nanoparticles \cite{sakiyama2004}.
Awasthi \textit{et al}. introduced a cohesive parameter which reduces the attractive interaction between atoms on the surface in their numerical model,
and simulate the rebound process of a Bi cluster onto a SiO$_2$ surface~\cite{awasthi2007}.
The qualitative validity of such a simplified model has been confirmed by the simulation of an atomic based model~\cite{saitoh2010}

Needless to say, fine powders play major roles in recent advanced nanotechnology and nanoscience.
Indeed, one of the main purposes of the nanotechnology and nanoscience is to understand, control and manipulate fine powders.
Because these fine powders in nanoscale are intermediate between single small molecules and macroscopic bulk materials, 
their properties and behavior is qualitatively different from those of their constituent elements and from those of macroscopic pieces of materials.
Therefore, it is important to understand the behavior of collisions of fine powders.

It was still believed that the restitution coefficient for normal head on collisions should be $e\le 1$ because this bound is connected with the second law of thermodynamics.
Nevertheless, it is remarkable that this bound is also violated even for normal head on collisions between nanoclusters, known as ``super rebound" for $e>1$, 
because thermal fluctuations can play a major role in nanoscale~\cite{kuninaka2009,han2010,kuninaka2012}.
%[
%Kuninaka09= H. Kuninaka and H. Hayakawa, Phys. Rev. E 79, 030309 (2009).
%Han10= L. B. Han, Q. An, S. N. Luo, W. A. Goddard, Material Lett. 20, 2033 (2010) 
%Kuninaka12=H. Kuninaka and H. Hayakawa, Phys. Rev. E 86, 051302 (2012)
%]
Motivated by these findings we only focus on normal head on collisions, though the tangential force plays important roles to describe rich behavior of cohesive collisions~\cite{dominik1997}.
In the super rebounds, parts of elastic vibrations are transferred to translational motion of the colliding bodies and thus kinetic energy of the translational motion of it can increase after the collision.
The super rebound is associated with the decreases of entropy~\cite{kuninaka2012} and the fluctuation theorem~\cite{evans1993, gallavotti1995, jarzynski1997, visco2005, chong2010, joubaud2012, naert2012, hayakawa2013}.
Indeed, Tasaki indicated that the probability of the restitution coefficient can satisfy an extended fluctuation theorem if the motion of the center of mass can be separated from the motion of internal degrees of freedoms~\cite{tasaki2006}.
%This is contrast to the conventional inelastic collisions in which kinetic energy is transferred into the vibration.
Note that the super rebounds take place only when the impact speed is nearly equal to or slower than the thermal speed to be consistent with the requirement of the thermodynamics.

The aim of this paper is to clarify the role of collective modes or visco-elastic vibrations inside the grain associated with the energy transfer between the translational motion and the internal modes.
For this purpose we extend the method developed for two-dimensional isothermal elastic disks~\cite{gerl1999,hayakawa2002} to three dimensional case.
It is remarkable that Aspelmeier performed a three dimensional simulation by introduction of an exponential potential $e^{- \alpha r}$ with the distance $r$ between atoms on the surface of a colliding sphere in the limit $\alpha \to \infty$.
Although his study is a pioneer work using a model of elastic spheres, his model produces some unnatural behavior such as the repeats of discrete contact and free flight during a collision and a harmonic contact force instead of expected Hertzian force~\cite{aspelmeier2000}.
%with the introduction of the attractive interaction on the surface of the colliding bodies or the wall.
To improve these points we propose a new model of colliding isothermal visco-elastic spheres as a natural extension of the previous 2D models~\cite{gerl1999, hayakawa2002}.

We also consider the effects of the solid viscosity, the attractive interaction on the surface of the colliding spheres or the wall and the initial temperature.
Most of the researches on the theory of elasticity assume that the local deformation takes place without dissipation. 
However this treatment is only valid for infinitesimal motion of local deformation. 
In real deformation taking place at finite speed violates the local force balance at each instance.
Thus, there exists a local relaxation process to recover the balance state, which causes the dissipation and the origin of irreversibly.
We only consider the dissipation associated with the local motion of atoms as in usual viscous fluids.

Our method is complementary to the method based on the molecular dynamics simulation~\cite{awasthi2006,suri2008,kuninaka2009,saitoh2010,tanaka2012,kuninaka2012},
in which we can know the detailed dynamics of constituent atoms in colliding objects but it is not appropriate to characterize the macroscopic deformation of the colliding objects.  
Of course, we cannot address structural phase transitions in the colliding nanoparticles unlike molecular dynamics simulations~\cite{valentini2007}, but such transitions only take place when the body collides at very high speed.
Indeed some studies based on the molecular dynamics simulation~\cite{kuninaka2009, tanaka2012} only observe elastic deformation in the colliding objects for slow impacts.
We also stress that the computational cost of our model is essentially independent of the cluster size in contrast to the molecular dynamics simulation.
Nevertheless, our model can consider the relevant effects of surface force which may play a major role in fine powders. 
%Therefore, we can easily compute the collision of the larger sphere than 10 nm.
%We address the macroscopic aspects of super rebounds.

The organization of this paper is as follows.
In the next section, we introduce our model of the colliding visco-elastic spheres and explain the set up of our simulation.
Section \ref{sec:athermal} is the results of our simulation at $T=0$, where we investigate the restitution coefficient against the impact speed.
It is remarkable that there exists an oscillation of the restitution coefficient against impact speed if the solid viscosity is absent or small, as will be shown in Sec. \ref{ssec:athermal_perfect}, though such an oscillation disappears for the large solid viscosity as in Sec. \ref{ssec:athermal_dissipative}.
We also investigate the excitation of vibrational modes against the contact duration to understand how the restitution coefficient depends on the impact speed.
In the third part (Sec. \ref{ssec:athermal_contact-mechanics}) we verify whether the conventional contact mechanics is reproducible for low speed collisions between spheres.
Section \ref{sec:thermal} exhibits the results of our simulation under the influence of thermal fluctuations.
In Sec. \ref{ssec:thermal_super} we study the mechanism of super rebounds ($e>1$) for collisions between a thermally activated cluster and a flat wall.
We also study the restitution coefficient against the initial phase of the vibration.
In Sec. \ref{ssec:thermal_fluctuation-theorem} we numerically confirm the fluctuation theorem for inelastic collisions introduced by Ref.~\cite{tasaki2006}.
In the third part (Sec. \ref{ssec:thermal_heating}) we confirm that the collisional heating during a collision is sufficiently small to be consistent with the assumption of our isothermal model.
In Sec. \ref{sec:discussion} we discuss our results. This section consists of three parts. 
In the first part (Sec. \ref{ssec:discussion_perturbation}) we develop the perturbation theory to explain the initial phase dependence of the restitution coefficient.
In the second part (Sec. \ref{ssec:discussion_mode-transfer}), we investigate the mode transfer when the initial condition contains only one mode excitation to clarify the mechanism of inelastic collisions. 
In the third part (Sec. \ref{ssec:discussion_perspective}), we summarize future work and perspectives.
% and the mechanism of excitation of quadrupole.
Finally, we summarize our conclusion in Sec. \ref{sec:conclusion}.
In Appendix \ref{sec:app_wave-equation}, we explain the derivation of the visco-elastic wave equation for isothermal spheres.
In Appendix \ref{sec:app_stressfree-solution}, we summarize the derivation of the stress free solutions of the wave equation.
In Appendix \ref{sec:app_fdr}, we explain the role of the fluctuating stress in continuum dynamics, and estimate the critical solid viscosity at which the relaxation time originated from the solid viscosity is comparable to the duration of contact.
In Appendices \ref{sec:app_force}, \ref{sec:app_relation-bri} and \ref{sec:app_perturbation-calc}, we summarize detailed calculations required for our model and results.

\section{\label{sec:model} MODEL}
In this section, we introduce our simulation model.
As stated in Introduction, we adopt a model of isothermal visco-elastic spheres.
The detailed derivation of the wave equation and its solution under the stress free boundary condition can be found in Appendices \ref{sec:app_wave-equation} and \ref{sec:app_stressfree-solution}, respectively, and the textbook by Love~\cite{love}.
%[Love=A. E. H.  Love, A treatise on the mathematical theory of elasticity (Dover, 1906).
Let us explain a general case in which a visco-elastic sphere is colliding on another visco-elastic sphere, at first.
%We also assume that the collision is normal head on, and there is no effect of the tangential interaction between the sphere and the wall.
Later, we will restrict our interest to the case of normal head on collisions between the sphere and a flat wall.

The equations of motion of two colliding visco-elastic spheres $i  = 1, 2$ for the radius $R_i$, mass density $\rho_i$ and mass $M_i \equiv \rho_i 4 \pi R_i^3 / 3$ are given by
\if0
\begin{alignat}{4} \label{eq:eom_com}
	M_\text{eff} \frac{\mathrm{d}^2 z_\text{CM}}{\mathrm{d} t^2} =
	&- \frac{\partial V(z_\text{CM}, \bm{u}_1, \bm{u}_2)}{\partial z_\text{CM}}, & \qquad & \\
	\label{eq:eom_displacement}
	\rho_i \frac{\partial^2 \bm{u}_i}{\partial t^2} =
	&\rho_i \left(1 + \gamma_i \frac{\partial}{\partial t} \right)
		\left\{ \left(c_i^{(\ell)} \right)^2 \bm{\nabla} \bm{\nabla} \cdot \bm{u}_i
		- \left(c_i^\text{(t)} \right)^2 \bm{\nabla} \times (\nabla \times \bm{u}_i) \right\} \nonumber & & \\
	&- \frac{\partial V(z_\text{CM}, \bm{u}_1, \bm{u}_2)}{\partial \bm{u}_i} & & (i = 1, 2),
\end{alignat}
\fi
\begin{eqnarray} \label{eq:eom_com}
	M_\text{eff} \ddot{z}_\text{CM} &=& - \frac{\partial V(z_\text{CM}, \bm{u}_1, \bm{u}_2)}{\partial z_\text{CM}}, \\
	\label{eq:eom_displacement}
	\rho_i \frac{\partial^2 \bm{u}_i}{\partial t^2} &=& \rho_i \left(1 + \gamma_i \frac{\partial}{\partial t} \right)
		\left\{ \left(c_i^{(\ell)} \right)^2 \bm{\nabla} \bm{\nabla} \cdot \bm{u}_i
		- \left(c_i^\text{(t)} \right)^2 \bm{\nabla} \times (\bm{\nabla} \times \bm{u}_i) \right\} \nonumber \\
	&& - \frac{\partial V(z_\text{CM}, \bm{u}_1, \bm{u}_2)}{\partial \bm{u}_i},
\end{eqnarray}
where $z_\text{CM}$ and $\bm{u}_i$ are, respectively, the distance between the centers of masses of the colliding two spheres and the deformation field.
%Here, we have introduced $\dot{\bm{u}}_i = \partial \bm{u}_i / \partial t$ and $\ddot{\bm{u}}_i = \partial^2 \bm{u}_i / \partial t^2$, respectively.
We also introduce the reduced mass $M_\text{eff} \equiv (1/M_1 + 1/M_2)^{-1}$, the longitudinal and the transverse sound speeds, $c_i^{(\ell)}$ and $c_i^\text{(t)}$ for the sphere $i$, and the solid viscosity $\gamma_i$.
We should note that the viscous terms proportional to $\gamma_i$ should be associated with the fluctuating stress to satisfy the fluctuation-dissipation relation or to relax in the equilibrium state (see Appendix \ref{sec:app_fdr}). Nevertheless, such the fluctuating stress is not important in our paper unless the case of large $\gamma_i$, because the duration time is much shorter than the equilibration time for most of our situations (see Appendix \ref{sec:app_fdr}). 
Therefore, we simply ignore the random noise terms in this paper.
Equation (\ref{eq:eom_displacement}) is the visco-elastic wave equation (see Appendix \ref{sec:app_wave-equation}) for the sphere $i$, where $V(z_\text{CM}, \bm{u}_1, \bm{u}_2)$ is the total interaction potential between atoms on the surface of the projectile and atoms on the surface of the target.
Here we assume that the velocity field $\bm{v}(t; \bm{x})$ of the elastic sphere is just the time derivative of the displacement: $\bm{v}(t; \bm{x}) = \partial \bm{u}(t; \bm{x}) / \partial t$.
Within this framework, the longitudinal and transversal solid viscosities are not independent (see Eq. (\ref{eq:viscosity_identity})).
We also assume the isothermal condition for the colliding spheres.
%, and therefore we do not need the corresponding thermal energy balance equation.
In Sec. \ref{ssec:thermal_heating}, we check that the assumption is self-consistent.

\begin{figure}
	\includegraphics[width = 86mm]{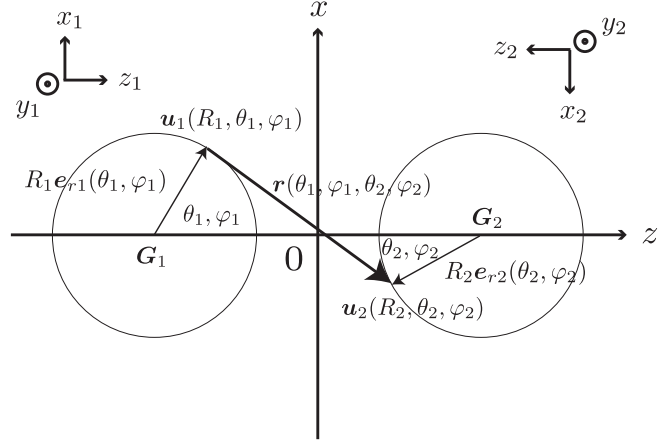}
	\caption{The coordinate system we adopt, where $\bm{u}_i(R_i, \theta_i, \varphi_i)$ represents the surface displacement of the cluster $i$.}
	\label{fig:coordinate_twospheres}
\end{figure}

We assume that Lennard-Jones atoms are distributed on the surface of the colliding sphere with planar density $d_i^{-2}$, where $d_i$ $(i = 1, 2)$ is the diameter of atoms placed on the sphere $i$.
The potential between the surface atoms at the distance $r$ is assumed to be described by
\begin{equation} \label{eq:lennard-jones}
	\phi(r) = 4 \epsilon \left[ \left( \frac{\sigma}{r} \right)^{12} - g \left( \frac{\sigma}{r} \right)^6 \right],
\end{equation}
where $\epsilon$ and $\sigma$ are the depth of the potential well and the diameter of the repulsive core, respectively, and we have introduced the cohesive parameter $0 \leq g \leq 1$ which reduces the attractive interaction~\cite{awasthi2007}.
Although we mainly investigate collisions without the reduction, $g = 1$, we also investigate the repulsive collision, $g = 0$ in Secs. \ref{ssec:athermal_contact-mechanics} and \ref{ssec:discussion_perturbation}, and the reduced attractive collision, $g = 0.2$ in Sec \ref{ssec:thermal_super}.
Here, the distance between the atoms on each surface can be written as (see Fig. \ref{fig:coordinate_twospheres})
\begin{eqnarray} \label{eq:distance}
	r(\theta_1, \varphi_1; \theta_2, \varphi_2) &=& |\bm{r}(\theta_1, \varphi_1; \theta_2, \varphi_2)| \nonumber \\
	&=& \left| \left\{ \bm{G}_2 + R_2 \bm{e}_{r2}(\theta_2, \varphi_2) + \bm{u}_2\left( R_2, \theta_2, \varphi_2 \right) \right\} \right. \nonumber \\
	& & \left. - \left\{ \bm{G}_1 + R_1 \bm{e}_{r1}(\theta_1, \varphi_1) + \bm{u}_1 \left(R_1, \theta_1, \varphi_1 \right) \right\} \right|,
\end{eqnarray}
where $\bm{G}_i$ and $\bm{e}_{ri}$ $(i = 1, 2)$ represent the position of center of mass and the radial unit vector for the sphere $i$, respectively.
Then, the total interaction energy is given by
\begin{equation} \label{eq:total_interaction_twospheres}
	V(z_\text{CM}, \bm{u}_1, \bm{u}_2) = \frac{R_1^2 R_2^2} {d_1^2 d_2^2}
		\int^{\pi/2}_0 \mathrm{d}\theta_1 \sin\theta_1 \int^{2\pi}_0 \mathrm{d}\varphi_1
		\int^{\pi/2}_0 \mathrm{d}\theta_2 \sin\theta_2 \int^{2\pi}_0 \mathrm{d}\varphi_2
		\phi(r(\theta_1, \varphi_1; \theta_2, \varphi_2)).
\end{equation}
When vibrational modes are not excited before the collision, the distance $r(\theta_1, \varphi_1; \theta_2, \varphi_2)$ depends only on $\theta_1$, $\theta_2$ and $\varphi_1 + \varphi_2$ for head-on collisions.
Then, the integration with respect to $\varphi_1$ and $\varphi_2$ in Eq. (\ref{eq:total_interaction_twospheres}) can be performed (see Appendix \ref{sec:app_force}).
On the other hand, when vibrational modes exist before the collision, we cannot execute such an integration even in head-on collisions.
If the second sphere (target) is a flat wall i.e. $R_2 \to \infty$ and the wall is also hard, i.e. $c_2^\text{(t)}$ and $c_2^{(\ell)}$ are infinite, the displacement of the wall $\bm{u}_2$ is identical to zero and the potential (\ref{eq:total_interaction_twospheres}) becomes a simpler form (see Eq. (\ref{eq:total_interaction_wall})).

Let us expand $\bm{u}_i$ in terms of a set of the dimensionless spheroidal modes for the cluster $i = 1, 2$ $\bm{\tilde{u}}_{i,n \ell m}^\text{(S)}(\bm{x})$ (see Eq. (\ref{eq:spheroidal-solution}))
% because this set is the complete orthonormal set \cite{}:
\begin{equation} \label{eq:expansion}
	\bm{u}_i(t; \bm{x}) = \sum_{n \ell m} Q_{i,n \ell m}(t) \tilde{\bm{u}}_{i,n \ell m}^\text{(S)}(\bm{x}),
\end{equation}
where $n$ $(n = 0, 1, 2, \ldots)$, $\ell$ $(\ell = 0, 1, 2, \ldots)$ and $m$ $(-\ell \leq m \leq \ell)$ are the radial, colatitudinal and azimuthal modes numbers, respectively.
We ignore the torsional modes (see Appendix \ref{sec:app_stressfree-solution}) because we restrict our interest to normal head-on collisions.
Substituting Eq. (\ref{eq:expansion}) into Eq. (\ref{eq:eom_displacement}), we obtain the equation of motion for the coefficient $Q_{i, n \ell m}(t)$:
\begin{equation} \label{eq:eom_vib}
	M_i \ddot{Q}_{i, n \ell m} = - M_i \omega_{i, n \ell}^2 (Q_{i, n \ell m} + \gamma_i \dot{Q}_{i, n \ell m})
		- \frac{\partial V(z_\text{CM}, \{Q_{i', n' \ell' m'}\})}{\partial Q_{i, n \ell m}},
\end{equation}
where $\omega_{i, n \ell}$ is the eigen-frequency of the sphere $i$ (see Appendix \ref{sec:app_stressfree-solution}).
We numerically solve Eqs. (\ref{eq:eom_com}) and (\ref{eq:eom_vib}) simultaneously.

For later discussion, let us introduce the eigen-energy $H_{i, n \ell m}(t)$:
\begin{equation}
	H_{i, n \ell m}(t) = \frac{1}{2} M_i \{\dot{Q}_{i, n \ell m}(t)\}^2 + \frac{1}{2} M_i \omega_{i, n \ell}^2 \{Q_{i, n \ell m}(t)\}^2,
\end{equation}
and the excitation energy $\Delta H_{i, n \ell m}$:
\begin{equation} \label{eq:excitation}
	\Delta H_{i, n \ell m} \equiv H_{i, n \ell m}(t_\text{f}) - H_{i, n \ell m}(0),
\end{equation}
where we have introduced duration time $t_\text{f}$ of the collision.
Then, the total energy $H_\text{tot}(t)$ of this system can be written by
\begin{equation}
	H_\text{tot}(t) = H_\text{CM}(t) + \sum_{i, n \ell m} H_{i, n \ell m}(t) + V(z_\text{CM}(t), \{Q_{1, n' \ell' m'}(t)\}, \{Q_{2, n' \ell' m'}(t)\}),
\end{equation}
where $H_\text{CM}(t)$ is the translational energy of the relative motion:
\begin{equation}
	H_\text{CM}(t) = \frac{1}{2} M_\text{eff} \{\dot{z}_\text{CM}(t)\}^2.
\end{equation}
%Introducing
%\begin{equation} \label{eq:total-vib-energy}
%	\Delta H_{i, \text{vib}} \equiv \sum_{n \ell m} \Delta H_{i, n \ell m}
%\end{equation}

In this paper, we mainly investigate collisions between a sphere and a hard wall, and collisions between homogeneous spheres (see TABLE \ref{table:summarize-situation}).
%: $c^\text{(t)} \equiv c_1^\text{(t)} = c_2^\text{(t)}$, $c^{(\ell)} \equiv c_1^{(\ell)} = c_2^{(\ell)}$, $\rho \equiv \rho_1 = \rho_2$ and $\gamma \equiv \gamma_1 = \gamma_2$
Even when we simulate collisions between two visco-elastic spheres, we assume that the spheres are made of identical atoms.
Thus, we can safely remove the subscript of the sphere $i$ for $c^\text{(t)}$, $c^{(\ell)}$, $\rho$ and $\gamma$ for later discussion.
%We take the radius $R_1$, the horizontal sound speed $c^\text{(t)}$ and the mass density $4\pi\rho/3$ as the units of this system.
To save the computational cost, we truncate the interaction potential at a cutoff distance $z_\text{cut} = 5 \sigma$, and we place the initial sphere out of interaction range.
We control the incident speed ranging from $0.001 c^\text{(t)}$ to $0.4 c^\text{(t)}$.
When we include the effects of the initial thermal fluctuations, we prepare the initial distributions of $Q_{i, n \ell m}(0)$ and $\dot{Q}_{i, n \ell m}(0)$ to satisfy the canonical distributions
\begin{eqnarray}
	\label{eq:canonical_q}
	p_\text{can}(Q_{i, n \ell m}(0)) &=& \sqrt{\frac{M_i \omega_{i, n \ell}^2}{2 \pi k_\text{B}T}}
		\exp \left[ - \frac{1}{k_\text{B}T} \frac{1}{2} M_i \omega_{i, n \ell}^2 \{Q_{i, n \ell m}(0) \}^2 \right], \\
	\label{eq:canonical_p}
	p_\text{can}(\dot{Q}_{i, n \ell m}(0)) &=& \sqrt{\frac{M_i}{2 \pi k_\text{B}T}}
		\exp \left[ - \frac{1}{k_\text{B}T} \frac{1}{2} M_i \{\dot{Q}_{n \ell m}(0) \}^2 \right], 
\end{eqnarray}
where we have introduced the temperature $T$ to characterize the variance of the initial fluctuations of modes in Eqs. (\ref{eq:canonical_q}) and (\ref{eq:canonical_p}).
When we are interested in collisions not affected by initial thermal fluctuations, we simply assume that there are no internal vibrations inside colliding spheres.
\begin{table}
	\caption{The parameter of the object $2$ for the collision between the colliding sphere and the flat wall and two spheres.}
	\begin{tabular}{c||cccc} \hline
	sphere and wall & $c_2^\text{(t)} \to \infty$ & $c_2^{(\ell)} \to \infty$ & ($R_2 \to \infty$) & \\ \hline
	two spheres & $c_2^\text{(t)} = c_1^\text{(t)}$ & $c_2^{(\ell)} = c_1^{(\ell)}$ & $\rho_2 = \rho_1$ & $\gamma_2 = \gamma_1$ \\ \hline
	\end{tabular}
	\label{table:summarize-situation}
\end{table}

\begin{figure}
	\begin{minipage}{0.5\columnwidth}\hspace{-\columnwidth}(a)
		\begin{center}
			\includegraphics[clip, width=\columnwidth]{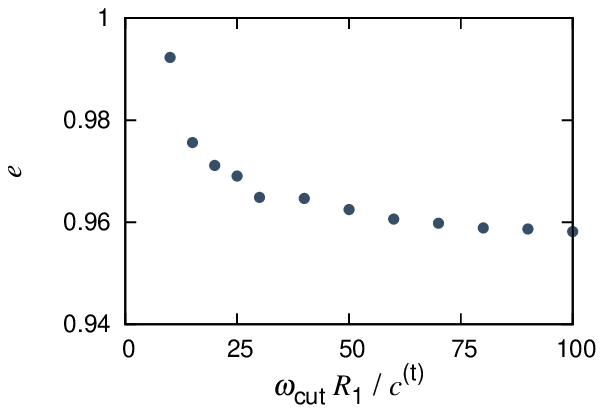}
		\end{center}
	\end{minipage}%
	\begin{minipage}{0.5\columnwidth}\hspace{-\columnwidth}(b)
		\begin{center}
			\includegraphics[clip, width=\columnwidth]{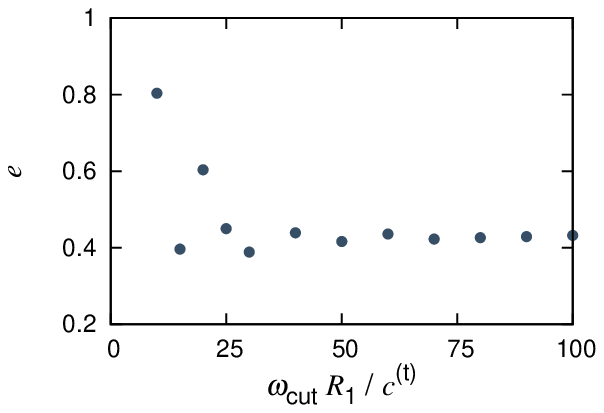}
		\end{center}
	\end{minipage}
	\caption{The cutoff frequency $\omega_\text{cut}$ dependence of the restitution coefficient for $T = 0$, $\gamma = 0$, and (a) $v_\text{CM}(0) = 0.1 c^\text{(t)}$ and (b) $v_\text{CM}(0) = 0.01 c^\text{(t)}$.}
%The restitution coefficient begins to converge at $\omega_\text{cut} = 50 c^\text{(t)} / R_1$ for (a) the faster impact $v_\text{CM}(0) = 0.1 c^\text{(t)}$ and $\omega_\text{cut} = 25 c^\text{(t)} / R_1$ for (b) the slower impact $v_\text{CM}(0) = 0.01 c^\text{(t)}$.}
	\label{fig:finite-mode-effect}
\end{figure}

We truncate the eigenmodes at the cutoff frequency $\omega_\text{cut}$ due to the limitation of our numerical resources.
Thus, the number of radial modes is determined from the condition $\omega_{i, n \ell} < \omega_\text{cut}$ for each $\ell$, where number of colatitudinal modes is approximately $\omega_\text{cut} R_1 / c^\text{(t)}$.
We adopt $\omega_\text{cut} = 100 c^\text{(t)} / R_1$ for the axisymmetric case, i.e. $T = 0$, in which the number of colatitudinal modes is $100$ and $m \equiv 0$, and we adopt $\omega_\text{cut} = 25 c^\text{(t)} / R_1$ for $T \neq 0$, in which the number of colatitudinal modes is $24$.
The total number of modes is approximately 1500.
Figure \ref{fig:finite-mode-effect} exhibits the convergence of the restitution coefficient against the cutoff frequency for $T = 0$, $\gamma = 0$.
The restitution coefficient begins to converge around at $\omega_\text{cut} = 25 c^\text{(t)} / R_1$ for both (a) the faster impact $v_\text{CM}(0) = 0.1 c^\text{(t)}$ and (b) the slower impact $v_\text{CM}(0) = 0.01 c^\text{(t)}$.
Thus, the numerical error due to the limitation of the mode number may be sufficiently small for our cutoff frequency.
%Thus, the number of modes we introduce is sufficiently large.

\begin{figure}
	\includegraphics[width = 86mm]{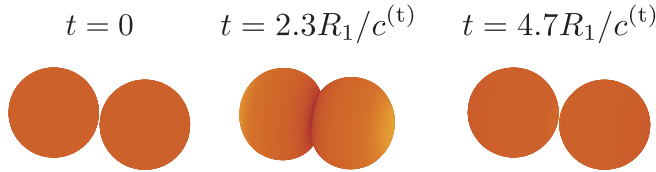}
	\caption{Time evolution of a collision between the identical elastic spheres for $v_\text{CM}(0) = 0.3 c^\text{(t)}$, $R_2 = R_1$, $T = 0$ and $\gamma = 0$, where the parameters we use are summarized in Table II. The middle figure corresponds to the moment of zero relative speed, in which the compression length of each cluster is approximately 20\%.}
	\label{fig:identical-spheres} 
\end{figure}

We adopt the Runge-Kutta-Fehlberg method with adaptive step for the integration of Eqs. (\ref{eq:eom_com}) and (\ref{eq:eom_vib}).
When there is no dissipation, i.e. $\gamma = 0$, the rate of energy conservation, $\left| H_\text{tot}(t) - H_\text{tot}(0) \right| / H_\text{tot}(0)$, is kept within $10^{-5}$.
%, where $H_\text{tot}(t) = H_\text{CM}(t) + H_{1, \text{vib}}(t) + H_{2, \text{vib}}(t) + V(z_\text{CM}(t), \{Q_{1, n' \ell' m'}(t)\}, \{Q_{2, n' \ell' m'}(t)\})$ is the total energy of this system.
We adopt Lebedev quadrature formula which is the Gaussian quadrature formula for the integration over the surface of a three-dimensional sphere~\cite{lebedev1975} to evaluate of the surface integral in Eq. (\ref{eq:total_interaction_twospheres})
\footnote{A Lebedev rule of precision $p$ can be used to correctly integrate any polynomial $f(x, y, z)$ for which the highest degree term $x^{a_x} y^{a_y} z^{a_z}$ satisfies $a_x + a_y + a_z \leq p$, where we adopt $p = 131$~\cite{lebedev1999}.}.
We adopt the simple trapezoidal rule to evaluate the interactive potential (\ref{eq:total_interaction_twospheres}) for the axisymmetric case.
To avoid unphysical setups, we use some parameters corresponding to the case of the copper for our simulation, which are summarized in TABLE \ref{table:param} \cite{agrawal2002}.
Figure \ref{fig:identical-spheres} is a series of snapshots of colliding two identical spheres to illustrate its time evolution.
The middle figure corresponds to the moment of zero relative speed, where the compression is approximately 20\%.
In this paper, we mainly investigate collisions for small spheres ($R_1 = 10$ nm), where we will discuss the sphere size dependence of collisions for $T = 0$ in Sec. \ref{ssec:athermal_perfect}.
\begin{table}\label{table:param}
	\caption{The copper's parameters we use in our simulation.}
	\begin{tabular}{cccccc} \hline
	$c^\text{(t)}$ & $\rho$ & Poisson's ratio & $\epsilon$ & $\sigma$ & $d$ \\ \hline
	2270 m/s & 8960 kg/m$^3$ & 0.343 & 0.415 eV & 0.2277 nm & 0.256 nm \\ \hline
	\end{tabular}
\end{table}

\section{\label{sec:athermal} Simulation at $T=0$}

In this section, we summarize the results of our simulation at $T=0$.
This section consists of three parts.
In Sec. \ref{ssec:athermal_perfect}, we show the results without the solid viscosity $\gamma=0$.
In Sec. \ref{ssec:athermal_dissipative}, we discuss the effects of $\gamma$.
In Sec. \ref{ssec:athermal_contact-mechanics}, we verify whether the contact mechanics of elasticity is held for slow impacts of elastic spheres.

\subsection{\label{ssec:athermal_perfect} The oscillation of the restitution coefficient for $\gamma=0$}

\begin{figure}
	\begin{minipage}{0.5\columnwidth}\hspace{-\columnwidth}(a)
		\begin{center}
			\includegraphics[clip, width=\columnwidth]{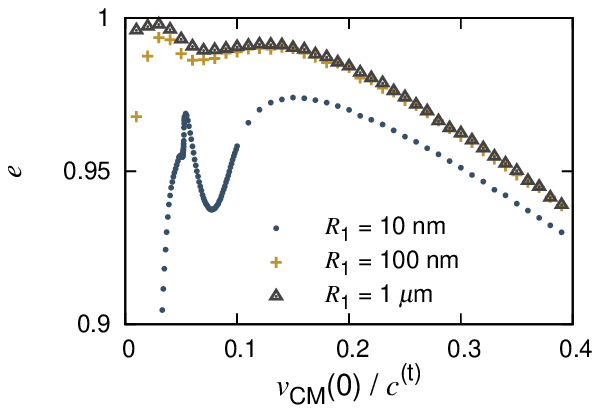}
		\end{center}
	\end{minipage}%
	\begin{minipage}{0.5\columnwidth}\hspace{-\columnwidth}(b)
		\begin{center}
			\includegraphics[clip, width=\columnwidth]{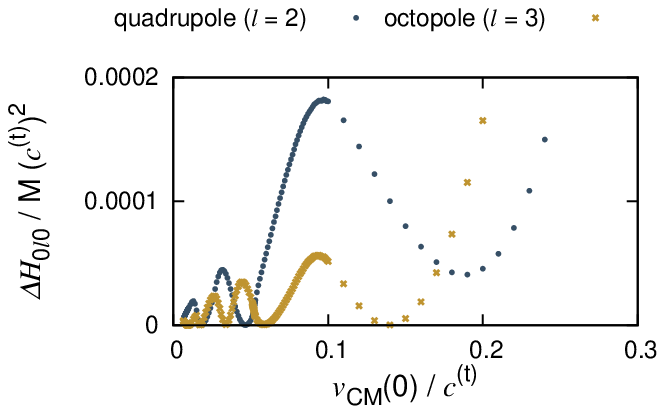}
		\end{center}
	\end{minipage}
	\caption{(a) The restitution coefficient as a function of the impact speed without dissipation, $\gamma = 0$ and (b) the excitation of the quadrupole ($\ell = 2$) and the octopole ($\ell = 3$) modes as a function of the impact speed.}
	\label{fig:v-dependence}
\end{figure}

We, here, investigate the impact speed dependence of the restitution coefficient for collisions between an elastic sphere ($\gamma = 0$) and a flat wall under an athermal initial condition at $T = 0$, ranging from $v_\text{CM}(0) = 0.001 c^\text{(t)}$ to $v_\text{CM}(0) = 0.4 c^\text{(t)}$.
Even without dissipation in our model, we can reproduce inelastic collisions characterized by $e < 1$ because the energy is transferred from the translational motion to vibrational modes during the contact.
From Fig. \ref{fig:v-dependence}(a), we find a characteristic oscillation of $e(v_\text{CM})$.
We plot the results for spheres of $R_1 = 10$ nm, $R_1 = 100$ nm and $R_1 = 1$ $\mu$m in Fig. \ref{fig:v-dependence}(a).
Because we adopt the radius $R_1$ as the length unit in our simulation, we practically change the parameters proportioned to $R_1$, such as the core diameter $\sigma = 0.02277 R_1$.
Here we discuss the size dependence of the collision.
The main difference between larger and smaller spheres is the strength of the surface attraction par the volume.
Indeed, the ``critical speed" below which the colliding spheres are coalesced for the smaller sphere is faster than that for the larger one.
%We also find that their restitution coefficients are nearly equal for faster impacts, in which the effect of the attraction on the collision is negligible.

We also investigate the excitation of each mode $\Delta H_{0 \ell 0}$ introduced in Eq. (\ref{eq:excitation}) against the impact speed, where we only focus on the fundamental modes ($n = 0$) because the excitations of the other modes are much smaller than that of the fundamental modes.
Figure \ref{fig:v-dependence}(b) shows the excitations of the quadrupole ($\ell = 2$) and the octopole ($\ell = 3$) modes against the impact speed.
We find the existence of regular oscillatory behavior in these relations.
%, in which the oscillatory period of the quadrupole mode is less than that of the octopole mode.
%There are some peaks with an almost identical interval in each excitation, which may suggest that the resonance between the collision and the vibration mode takes place.

\begin{figure}
	\begin{minipage}{0.5\columnwidth}\hspace{-\columnwidth}(a)
		\begin{center}
			\includegraphics[clip, width=\columnwidth]{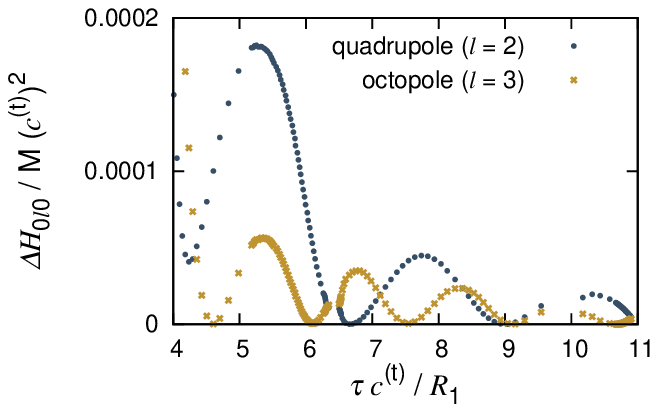}
		\end{center}
	\end{minipage}%
	\begin{minipage}{0.5\columnwidth}\hspace{-\columnwidth}(b)
		\begin{center}
			\includegraphics[clip, width=\columnwidth]{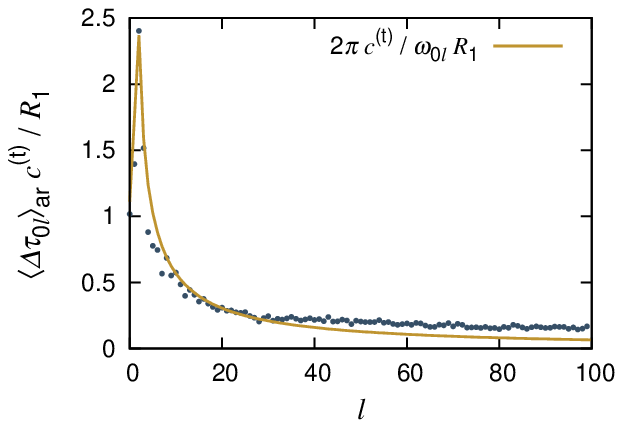}
		\end{center}
	\end{minipage}
	\caption{(a) The excitation of the quadrupole and the octopole modes against the contact duration and (b) the arithmetic mean of the intervals $\langle \Delta \tau_{0 \ell} \rangle_\text{ar}$ between the local minimums of $\Delta H_{0l0}(\tau)$ for all the fundamental modes. The solid line in (b) represents the inverse of eigen-frequencies $2\pi/\omega_{0 \ell}$.}
	\label{fig:resonance}
\end{figure}

Then, we replace the impact speed by the contact duration $\tau$ which is only the time scale except for the eigen-frequencies in this system (see Fig. \ref{fig:resonance}(a)).
Here we introduce the potential cutoff $V_\text{cut}$ to suppress the long-ranged tail effect in the interactions of slow impacts as $V(z_\text{CM}, \{Q_{n \ell 0}\}) = 0$ if the calculated potential is smaller than $V_\text{cut}$.
%%Note that we take $V_\text{cut}$ greater than the total interaction at the cutoff distance, $V(z_\text{cut}, \{Q_{n \ell 0}\})$ to remove the long time tail of the interaction.
Figure \ref{fig:resonance}(a) shows the relation between the excitation $\Delta H_{0 \ell 0}$ and the contact duration $\tau$ for $V_\text{cut} = 5 \times 10^{-5} M (c^\text{(t)})^2$.
We find that the oscillation period of $\Delta H_{0 \ell 0} (\tau)$ is a constant, where the period for $\ell = 2$ is larger than the period for $\ell = 3$.
The oscillatory behavior is supposed to be caused by the resonance between the duration of contact and the oscillation period of vibration of each mode.
%If the appearance of the peaks is caused by the resonance between the collision and the vibration of each mode, the equivalent interval $\Delta \tau_{0 \ell}$ should correspond to the inverse of the eigen-frequency $2\pi/\omega_{0 \ell}$.
To confirm this conjecture, we evaluate the arithmetic mean of the intervals $\langle \Delta \tau_{0 \ell} \rangle_\text{ar}$ between the local minimums of $\Delta H_{0 \ell 0} (\tau)$, and compare it with $2\pi/\omega_{0 \ell}$ for all the fundamental modes as plotted in Fig. \ref{fig:resonance}(b).
They are in good agreement with each other except for cases with very large $\ell (\geq 30)$, and thus we can conclude that oscillations of $e(v_\text{CM})$ and $\Delta H_{0 \ell 0}(v_\text{CM})$ are caused by the resonance between the duration of contact and the vibration period of each mode.
The difference between $\langle \Delta \tau_{0 \ell} \rangle_\text{ar}$ and $2\pi/\omega_{0\ell}$ for larger $\ell$ may be originated from the limitation of the resolution of the duration, which is approximately $0.1 R_1 / c^\text{(t)}$ (see Fig. \ref{fig:resonance}(a)).

\subsection{\label{ssec:athermal_dissipative} The restitution coefficient for finite solid viscosity}

\begin{figure}
	\begin{minipage}{0.5\columnwidth}\hspace{-\columnwidth}(a)
		\begin{center}
			\includegraphics[clip, width=\columnwidth]{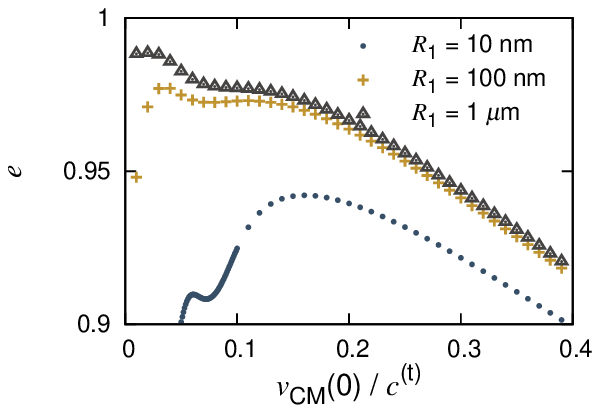}
		\end{center}
	\end{minipage}%
	\begin{minipage}{0.5\columnwidth}\hspace{-\columnwidth}(b)
		\begin{center}
			\includegraphics[clip, width=\columnwidth]{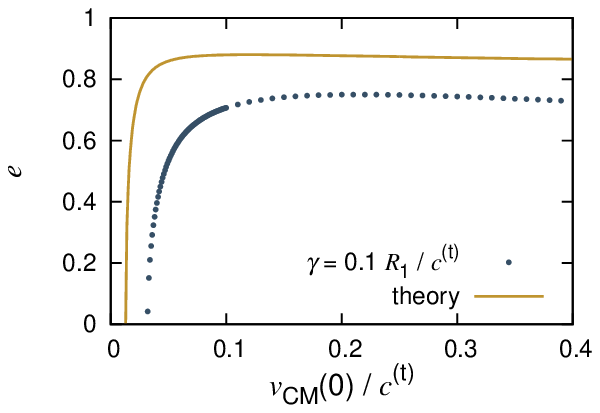}
		\end{center}
	\end{minipage}
	\caption{The restitution coefficient against the impact speed for (a) $\gamma = 0.01 R_1 / c^\text{(t)}$ and (b) $\gamma = 0.1 R_1 / c^\text{(t)}$, respectively. The solid line in (b) represents the theoretical prediction of cohesive collisions between visco-elastic spheres.}
	\label{fig:e-v_dissipation}
\end{figure}

In this subsection, we study collisions of visco-elastic spheres for finite $\gamma$ at $T=0$. 
Figure \ref{fig:e-v_dissipation} exhibits the results of $e(v_{CM})$ for finite $\gamma$, where the oscillation of $e(v_\text{CM})$ still remains for $\gamma = 0.01 R_1 / c^\text{(t)}$, whereas it disappears for $\gamma = 0.1 R_1 / c^\text{(t)}$.
It should be noted that the behavior of the restitution coefficient for small $\gamma$ is quite different from known results from the quasi-static theory, but its behavior for $\gamma = 0.1 R_1 / c^\text{(t)}$ is similar to the known one. 
Indeed, the solid line in Fig. \ref{fig:e-v_dissipation}(b) represents the theoretical prediction of cohesive collisions between visco-elastic spheres~\cite{brilliantov2007}, where the force between cohesive spheres is described as the function of the contact radius $a$ and the speed $\dot{a}$ as
\begin{equation} \label{eq:bri-force}
	F(a, \dot{a}) = \frac{4 Y_\text{eff} a^3}{3 R_\text{eff}} - \sqrt{8 \pi Y_\text{eff} G} a^{3/2} + \gamma \dot{a} \frac{\partial}{\partial a} F(a, \dot{a}),
\end{equation}
where 
\begin{eqnarray}
	R_\text{eff} &\equiv& \left( \frac{1}{R_1} + \frac{1}{R_2} \right)^{-1}, \\
	Y_\text{eff} &\equiv& \left( \frac{1 - \nu_1^2}{Y_1} + \frac{1 - \nu_2^2}{Y_2} \right)^{-1},
\end{eqnarray}
are, respectively, the reduced radius and effective Young's modulus of two spheres with Young moduli $Y_1, Y_2$ and Poisson's ratios $\nu_1, \nu_2$.
$G$ in Eq. (\ref{eq:bri-force}) is the surface tension satisfying $G = 25 \pi \epsilon \sigma^4 / 24 d^6$~\cite{israelachvili2011, kuninaka2009}.
We note that the coefficient $\gamma$ in the third term on the right hand side of Eq. (\ref{eq:bri-force}) is identical to that used by Brilliantov \textit{et al}.~\cite{brilliantov2007} if there exists only one solid viscosity $\gamma$ for collisions between spheres of identical constituents (see Appendix \ref{sec:app_relation-bri}).
We numerically solve the equation of motion with the force (\ref{eq:bri-force}) using the fourth order Runge-Kutta method with adaptive time interval and plot the solid line in Fig. \ref{fig:e-v_dissipation}(b) without any fitting parameter.
It is easily verified that the theory reproduces the qualitative behavior of the restitution coefficient, but there is no quantitative agreement with the simulation for $\gamma = 0.1 R_1 / c^\text{(t)}$.
So far we do not identify the reason why we have large dissipation in the simulation.
One of the possibilities is that the model we use, Eq. (\ref{eq:eom_displacement}), may not correspond to the quasi-static model.
Indeed, there is a vibrational excitation in our model in addition to the solid viscosity for the dissipation mechanism.
Another possibility is that the deviation may come from the neglect of the fluctuating stress at finite temperature introduced in Appendix \ref{sec:app_fdr}.
As mentioned in Appendix, the fluctuating stress may play a role for large $\gamma$, but we simply ignore its role.

\subsection{\label{ssec:athermal_contact-mechanics} The contact mechanices}

\begin{figure}
	\begin{minipage}{0.5\columnwidth}\hspace{-\columnwidth}(a)
		\begin{center}
			\includegraphics[clip, width=\columnwidth]{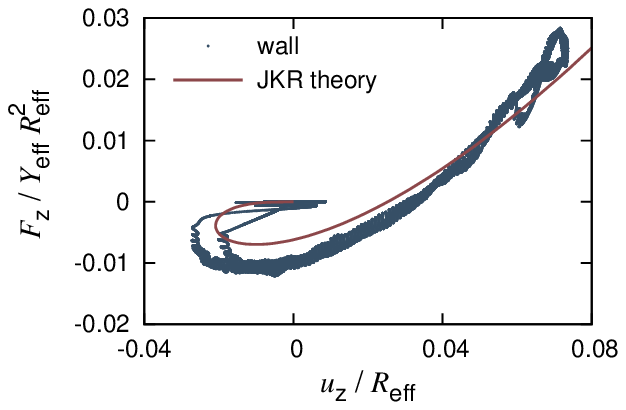}
		\end{center}
	\end{minipage}%
	\begin{minipage}{0.5\columnwidth}\hspace{-\columnwidth}(b)
		\begin{center}
			\includegraphics[clip, width=\columnwidth]{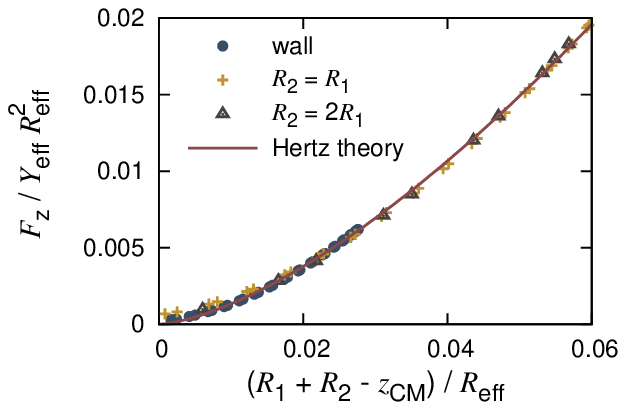}
		\end{center}
	\end{minipage}
	\caption{The scaled force as a function of the displacement at the pole $(r, \theta, \phi) = (R_1, 0, 0)$ without the solid viscosity for (a) the attractive ($g = 1$) and (b) the repulsive ($g = 0$) cases, respectively. Here, ``wall" represents the results of simulation for collision between the sphere and the wall.}
	\label{fig:contact-mechanics}
\end{figure}

In this subsection we investigate the deformation of elastic spheres under an applied force $F_z \equiv - \partial V / \partial z_\text{CM}$ during slow impacts, $v_\text{CM}(0) = 0.01 c^\text{(t)}$, at the pole $(r, \theta, \phi) = (R_1, 0, 0)$ ($\gamma = 0$).
First we study the case that the attractive force between the spheres exists.
The solid line in Fig. \ref{fig:contact-mechanics}(a) represents the prediction of JKR theory~\cite{johnson1971, johnson1985}.
Here, the force $F_z$ in the vertical line is scaled by the reduced radius $R_\text{eff}$ and Young's modulus $Y_\text{eff}$.
When we simulate collisions between an isothermal elastic sphere and the wall with the cohesive parameter $g = 1$,
we found an interesting hysteresis loop in the contact force as reported by Tanaka \textit{et al}.~\cite{tanaka2012}.
The time evolution of the contact force is almost reproducible by the JKR theory~\cite{johnson1971, johnson1985}, though the theory cannot reproduce the hysteresis loop as reported in MD~\cite{tanaka2012}.
In summary, we verify the relevancy of the theory of elasticity even for quasi-static contact process.

Next we study the repulsive case for $g = 0$ (Fig. \ref{fig:contact-mechanics}(b)).
So far we have mainly simulated collisions between an isothermal elastic sphere and a flat wall.
We also simulate collisions between two isothermal elastic spheres with different radii and identical Young's moduli and Poisson's ratios to verify the validity of the theory of elasticity in the quasi-static region~\cite{hertz1882, landau1959}.
%We explain the coordinate system of particle-particle collisions in Appendix \ref{sec:app_coordinate-withparticle}.
Figure \ref{fig:contact-mechanics}(b) plots the scaled force $F_z / Y_\text{eff} \sqrt{R_\text{eff}}$ for collisions between an elastic sphere with the radius $R_1$ and the wall ($R_2 \to \infty$), between an identical sphere with $R_2 = R_1$, and between a sphere with $R_1$ and a sphere with $R_2 = 2 R_1$ for  $v_\text{CM}(0) = 0.01 c^\text{(t)}$.
Here we plot $R_1 + R_2 - z_\text{CM}$ ($R_1 - z_\text{CM}$ for the collision with the wall) as the horizontal coordinate, while we plot the displacement $u_z(R_1, 0, 0)$ for the attractive case.
As expected from the theory of elasticity, we verify that the scaled force $F_z / Y_\text{eff} \sqrt{R_\text{eff}}$ for the three dimensional simulation can reproduce the contact theory which is independent of the target radius $R_2$, though the value of the horizontal coordinate needs a shift of the origin because we use the soft potential, where the solid line in Fig. \ref{fig:contact-mechanics}(b) predicted by the Hertz theory~\cite{landau1959} is perfectly on the simulation data only shift of the origin in horizontal coordinate.

\section{\label{sec:thermal} Collision at finite $T$}

In this section, we study collisions under the influence of thermal fluctuations ($T\ne 0$). 
Here we restrict our interest to collisions of the small sphere ($R_1 = 10$ nm) which is strongly fluctuated at finite temperature against the wall.
This section consists of three parts. 
In Sec. \ref{ssec:thermal_super}, we discuss the mechanism of super rebounds, $e>1$ in details. 
In Sec. \ref{ssec:thermal_fluctuation-theorem}, we verify the existence of the extended fluctuation theorem proposed by Tasaki~\cite{tasaki2006}. 
In Sec. \ref{ssec:thermal_heating}, we confirm the theoretical consistency in which the heating induced by collisions is sufficiently small to be consistent with the isothermal elastic model.  

\subsection{\label{ssec:thermal_super} Super rebounds}

\begin{figure}
	\begin{minipage}{0.5\columnwidth}\hspace{-\columnwidth}(a)
		\begin{center}
			\includegraphics[clip, width=\columnwidth]{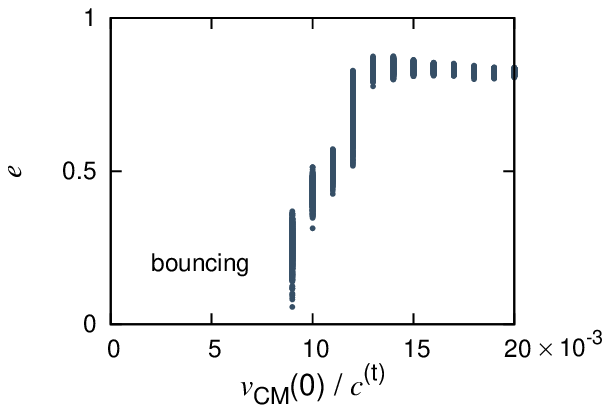}
		\end{center}
	\end{minipage}%
	\begin{minipage}{0.5\columnwidth}\hspace{-\columnwidth}(b)
		\begin{center}
			\includegraphics[clip, width=\columnwidth]{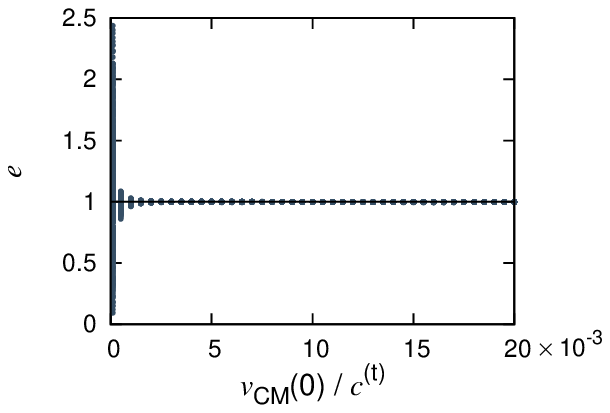}
		\end{center}
	\end{minipage}
	\caption{The impact speed dependence of the restitution coefficient for $T = 2.14 \times 10^{-8} M (c^\text{(t)})^2$ case with fixing (a) $g = 1$ and (b) $g = 0.2$, respectively. The temperature corresponds to 300 K in the physical unit.}
	\label{fig:e-v_temp300K}
\end{figure}

In this subsection, we study the mechanism of super rebounds where the restitution coefficient exceeds unity.
We take 1000 samples for each impact in Fig. \ref{fig:e-v_temp300K} at $T = 2.14 \times 10^{-8} M (c^\text{(t)})^2$ (300 K in the physical unit).
We only observe either conventional inelastic collisions i.e. $e<1$ or coalescence for $g = 1$, whereas we observe super rebounds $e > 1$ for the suppressed attraction case, $g = 0.2$.

\begin{figure}
	\includegraphics[width = 86mm]{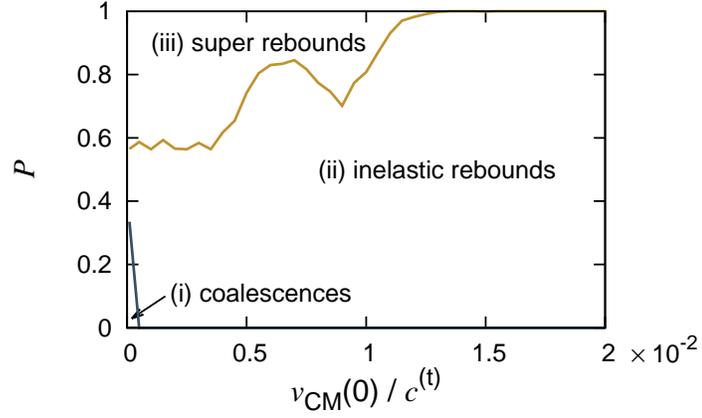}
	\caption{Probability diagrams classified by the three collision modes for $g = 0.2$. The regions (i), (ii) and (iii) represent coalescences, ordinary inelastic rebounds and super rebounds, respectively.}
	\label{fig:p-v} 
\end{figure}

We investigate the emergence probability of three modes in the collisions: (i) bouncing, (ii) normal inelastic collision for $e < 1$, and (iii) super rebounds for $e > 1$.
Figure \ref{fig:p-v} shows the phase diagram which is obtained under the fixed cohesive parameter $g = 0.2$, where $P$ represents the probability to observe each mode.
We take 1000 samples to evaluate $P$.
This phase diagram exhibits that the regions for the bouncing (i) decrease with the increase of the impact speed.
The super rebounds can be observed within the range of impact speed $v_\text{CM}(0) \leq 0.013 c^\text{(t)}$.
In addition, the probability to appear the super rebounds has a peak at $v_\text{CM}(0) = 0.009 c^\text{(t)}$ due to the resonance with eigenmodes.

\begin{figure}
	\includegraphics[width = 86mm]{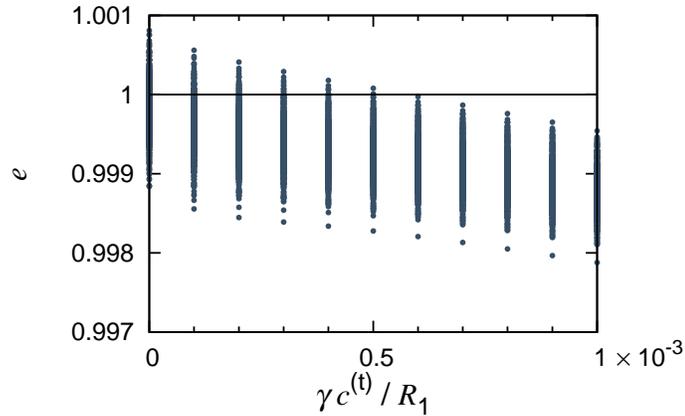}
	\caption{The restitution coefficient against the solid viscosity $\gamma$ for the impact speed $v_\text{CM}(0) = 0.009 c^\text{(t)}$.}
	\label{fig:e-diss}
\end{figure}

Figure \ref{fig:e-diss} shows the relation between the restitution coefficient and the solid viscosity $\gamma$ for the impact speed $v_\text{CM}(0) = 0.009 c^\text{(t)}$.
We find that the events of super rebounds decreases as $\gamma$ increases, and disappears for $\gamma > 6 \times 10^{-4} R_1 / c^\text{(t)}$.

\begin{figure}
	\begin{minipage}{0.5\columnwidth}\hspace{-\columnwidth}(a)
		\begin{center}
			\includegraphics[clip, width=\columnwidth]{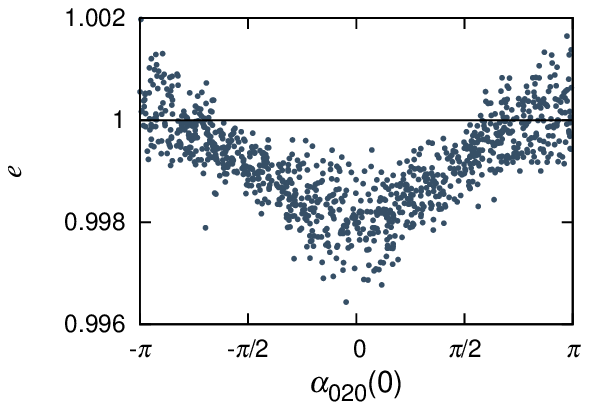}%
		\end{center}
	\end{minipage}%
	\begin{minipage}{0.5\columnwidth}\hspace{-\columnwidth}(a$'$)
		\begin{center}
			\includegraphics[clip, width=\columnwidth]{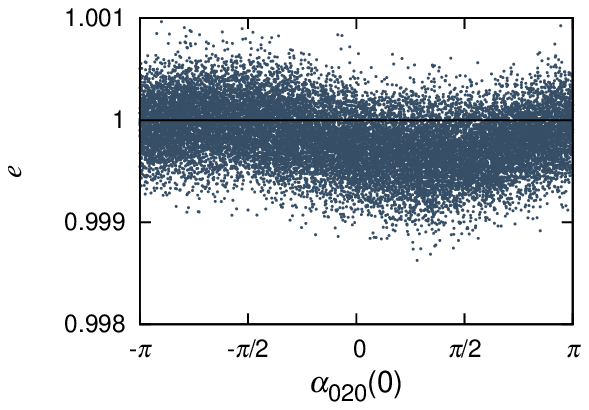}%
		\end{center}
	\end{minipage}

	\begin{minipage}{0.5\columnwidth}\hspace{-\columnwidth}(b)
		\begin{center}
			\includegraphics[clip, width=\columnwidth]{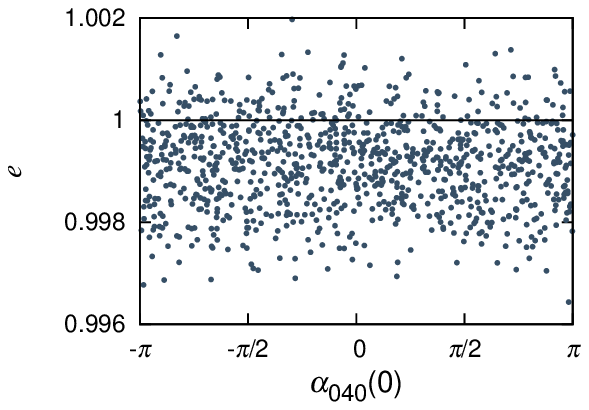}%
		\end{center}
	\end{minipage}%
	\begin{minipage}{0.5\columnwidth}\hspace{-\columnwidth}(b$'$)
		\begin{center}
			\includegraphics[clip, width=\columnwidth]{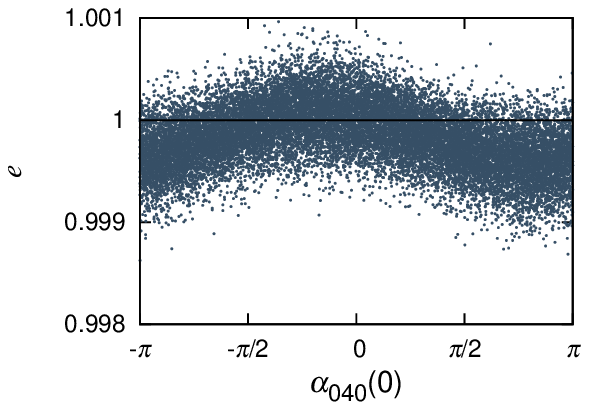}%
		\end{center}
	\end{minipage}

	\begin{minipage}{0.5\columnwidth}\hspace{-\columnwidth}(c)
		\begin{center}
			\includegraphics[clip, width=\columnwidth]{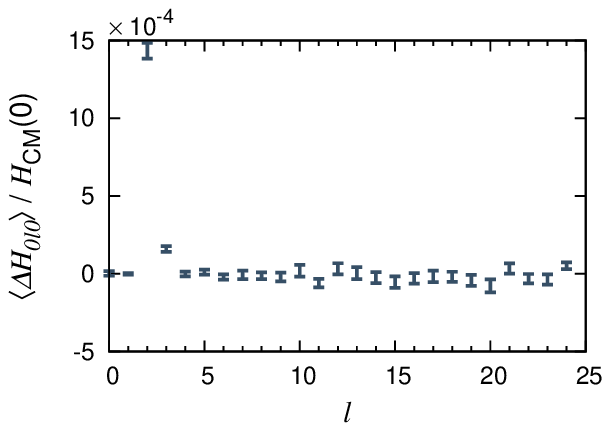}%
		\end{center}
	\end{minipage}%
	\begin{minipage}{0.5\columnwidth}\hspace{-\columnwidth}(c$'$)
		\begin{center}
			\includegraphics[clip, width=\columnwidth]{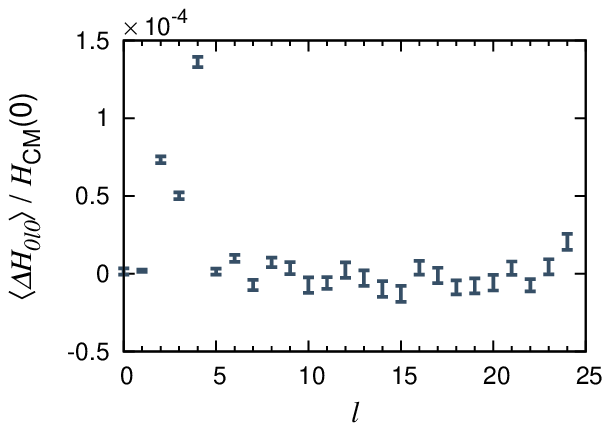}%
		\end{center}
	\end{minipage}
	\caption{The relation between the restitution coefficient and the initial phase of (a), (a$'$) the quadrupole ($\ell = 2$) mode and (b), (b$'$) the 16-pole ($\ell = 4$) mode. We plot in (c) and (c$'$) the excitation of each fundamental ($n = 0$) and axial ($m = 0$) eigenmode. (a), (b) and (c) are the results for $v_\text{CM}(0) = 0.007 c^\text{(t)}$ and include 1000 samples, whereas (a$'$), (b$'$) and (c$'$) are the results for $v_\text{CM}(0) = 0.009 c^\text{(t)}$ and include 20000 samples. The error bar in (c) and (c$'$) represents the standard deviation.}
	\label{fig:super_mode}
\end{figure}

%		\input{../321results_super_mode.tex}
%Results
%Super rebound
Here we focus on samples for $v_\text{CM}(0) = 0.007 c^\text{(t)}$, in which the probability of super rebounds becomes local minimum against the impact speed (see Fig. \ref{fig:p-v}), and samples for $v_\text{CM}(0) = 0.009 c^\text{(t)}$ to clarify the mechanism of super rebounds.
In Fig. \ref{fig:super_mode}, (a), (b) and (c) correspond to the results for $v_\text{CM}(0) = 0.007 c^\text{(t)}$, while (a$'$), (b$'$) and (c$'$) correspond to the results for $v_\text{CM}(0) = 0.009 c^\text{(t)}$.
Figure \ref{fig:super_mode}(a), (a$'$), (b) and (b$'$) exhibit the relations between the restitution coefficient and the initial phase of either the quadrupole ($\ell = 2$) mode or the 16-pole ($\ell = 4$) mode.
%We also investigate the relation between the restitution coefficient and the initial phase of each eigenmode.
Here the initial phase $\alpha_{n \ell m}$ is determined by
\begin{equation}
	Q_{n \ell m}(0) = \frac{1}{\omega_{n \ell}} \sqrt{\frac{2 H_{n \ell m}(0)}{M}} \sin \alpha_{n \ell m}(0),
\end{equation}
We find the sinusoidal structure of the restitution coefficient in Fig. \ref{fig:super_mode}(a), (a$'$) and (b$'$), whereas Fig. \ref{fig:super_mode}(b) displays the uniform distribution.
In particular, the curve for $v_\text{CM}(0) = 0.007 c^\text{(t)}$ and $\ell = 2$ (Fig. \ref{fig:super_mode}(a)) has the very large amplitude.
These results suggest that the initial phases for some modes play key roles to generate super rebounds.
We also investigate the excitation energy of each mode with the collision.
Figure \ref{fig:super_mode}(c) and (c$'$) show the averaged excitation energy $\langle \Delta H_{0 \ell 0} \rangle$ of the fundamental ($n = 0$) and the axial ($m = 0$) modes scaled by the initial kinetic energy $H_\text{CM}(0) = M \{v_\text{CM}(0)\}^2 / 2$, where the error bar in these figures represents the standard deviation.
The quadruple ($\ell = 2$) mode is strongly excited for $v_\text{CM}(0) = 0.007 c^\text{(t)}$, whereas its excitation is suppressed and the 16-pole ($\ell = 4$) mode is most excited for $v_\text{CM}(0) = 0.009 c^\text{(t)}$.
The excitation energy seems to be correlated with the amplitude of the sinusoidal curve.
Indeed, we will examine the perturbation theory of Eqs. (\ref{eq:eom_com}) and (\ref{eq:eom_vib}) to clarify the mutual relationship in Sec. \ref{ssec:discussion_perturbation}.
It should be noted that the excitation of the quadrupole mode for $v_\text{CM}(0) = 0.009 c^\text{(t)}$ is approximately 20 times smaller than the excitation for $v_\text{CM}(0) = 0.007 c^\text{(t)}$ in spite of the faster collision.
%The excitation of the modes is strongly affected by the relation between the impact speed or the contact duration and the eigen-frequency as we discuss in Sec \ref{ssec:athermal_perfect} (Fig. \ref{fig:v-dependence}).
%The impact speed dependence of the excitation has been discussed in Sec \ref{ssec:athermal_perfect} (Fig. \ref{fig:v-dependence}).
The quadrupole mode is the lowest order mode.
% and the deformation of the sphere caused by its oscillation is similar to the deformation with normal head-on collisions.
Thus, the quadrupole may be most strongly excited unless the resonance between the collision and the oscillation takes place.
The large suppression of the quadrupole excitation may also cause the probability of super rebounds to be large at $v_\text{CM}(0) = 0.009 c^\text{(t)}$.
%The horizontal coordinate expresses the number to characterize a complete set of numbers to specify eigen-modes $(n, l, m)$:
%$0 = (0, 0, 0), 1 = (1, 0, 0), 2 = (2, 0, 0), 3 = (0, 1, -1), 4 = (0, 1, 0), 5 = (0, 1, 1), 6 = (1, 1, -1), \dotsc, 1553 = (0, 24, -23), 1554 = (0, 24, -24).$
\subsection{\label{ssec:thermal_fluctuation-theorem} Fluctuation Theorem}

Fluctuation theorem states that the ratio of the probability of positive entropy production to the probability of negative entropy production can be expressed by an exponential function in systems out of equilibrium~\cite{evans1993, evans2002}.
Under an assumption of separation between the macroscopic translational mode and the microscopic internal modes, Tasaki extended the fluctuation theorem to the case of inelastic collisions:~\cite{tasaki2006}:
\begin{equation} \label{eq:fluctuation_theorem}
	\frac{P(X_0 \to X_1)}{P(\overline{X}_1 \to \overline{X}_0)} = e^{- W(X_0 \to X_1) / k_\text{B} T} 
\end{equation}
where $X_0 \equiv (z_\text{CM}(0), v_\text{CM}(0))$ and $X_1 \equiv (z_\text{CM}(t_f), v_\text{CM}(t_f))$ are the macroscopic variables at initial and final states, respectively, while $\overline{X}_0 = (z_\text{CM}(0), - v_\text{CM}(0))$ and $\overline{X}_1 = (z_\text{CM}(t_f), - v_\text{CM}(t_f))$ are the states obtained by reversing all the velocity in $X_0$ and $X_1$, respectively.
Here, $P(X_0 \to X_1) \mathrm{d}X_1$ is the transition probability of the macroscopic states from fixed $X_0$ into the interval between $X_1$ and $X_1 + \mathrm{d}X_1$ and $P(\overline{X}_1 \to \overline{X}_0) \mathrm{d}\overline{X}_0$ is the transition probability from fixed $\overline{X}_1$ into the interval between $\overline{X}_0$ and $\overline{X}_0 + \mathrm{d}\overline{X}_0$, and $W(X_0 \to X_1) \equiv M [\{v_\text{CM}(t_\text{f})\}^2 - \{v_\text{CM}(0)\}^2] / 2$ is the macroscopic energy loss during the transition from $X_0$ to $X_1$.
If $W(X_0 \to X_1) > 0$, $P(X_0 \to X_1) \mathrm{d}X_1$ is the probability of super rebounds, which is exponentially small probability of the ordinary inelastic collisions $P(\overline{X}_1 \to \overline{X}_0) \mathrm{d}\overline{X}_0$.
Although Kuninaka and Hayakawa~\cite{kuninaka2009_2} examined whether the fluctuation theorem is valid for inelastic collisions based on their molecular dynamics simulation, their result does not support the existence of the fluctuation theorem.

\begin{figure}
	\begin{minipage}{0.5\columnwidth}\hspace{-\columnwidth}(a)
		\begin{center}
			\includegraphics[clip, width=\columnwidth]{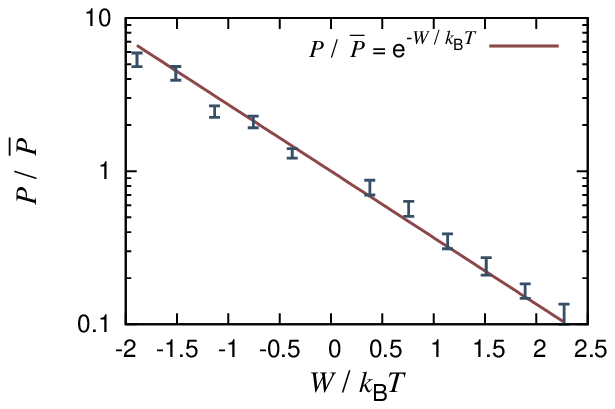}
		\end{center}
	\end{minipage}%
	\begin{minipage}{0.5\columnwidth}\hspace{-\columnwidth}(b)
		\begin{center}
			\includegraphics[clip, width=\columnwidth]{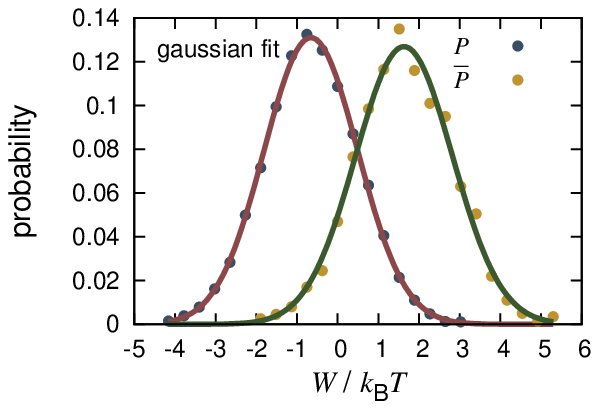}
		\end{center}
	\end{minipage}
	\caption{\label{fig:fluctuation_theorem} (a) The relation between $P / \overline{P}$ and $W$ at $T = 2.14 \times 10^{-8} M (c^\text{(t)})^2$ (300 K in the physical unit) and $v_\text{CM}(0) = 0.009 c^\text{(t)}$, and (b) the probability distributions $P$ and $\overline{P}$ against $W / k_\text{B} T$, in which the initial speeds are $0.009 c^\text{(t)}$ and $0.0090054 c^\text{(t)}$, respectively.}
\end{figure}

Figure \ref{fig:fluctuation_theorem}(a) shows the ratio of time normal to reversal probability distributions $P / \overline{P}$ against the macroscopic energy loss $W(X_0 \to X_1)$ observed in our simulation for $g = 0.2$, where we define $P \equiv P(X_0 \to X_1)$ and $\overline{P} \equiv P(\overline{X}_1 \to \overline{X}_0)$.
We take $N_\text{tot} = 20000$ samples at $T = 2.14 \times 10^{-8} M (c^\text{(t)})^2$ (300 K in the physical unit) and $v_\text{CM}(0) = 0.009 c^\text{(t)}$, whereas we take $\overline{N}_\text{tot}$ samples for various initial speeds.
$\overline{N}_\text{tot}$ is larger than 1000, while $\overline{N}_\text{tot}$ depends on $\overline{X}_1$
We evaluate probabilities $P$ and $\overline{P}$ as
\begin{equation}
	P = \frac{N_\text{eve}}{N_\text{tot}}, \qquad \overline{P} = \frac{\overline{N}_\text{eve}}{\overline{N}_\text{tot}},
\end{equation}
where $N_\text{eve}$ and $\overline{N}_\text{eve}$ are the numbers of events of the transition from fixed $v_\text{CM}(0) = 0.009 c^\text{(t)}$ into the interval between $v_\text{CM}(t_f) - \Delta v / 2$ and $v_\text{CM}(t_f) + \Delta v / 2$ and the transition from fixed $- v_\text{CM}(t_f)$ into the interval between $- v_\text{CM}(0) - \Delta v / 2$ and $- v_\text{CM}(0) + \Delta v / 2$, respectively.
We adopt the bin width $\Delta v = 10^{-4} v_\text{CM}(0)$, and $N_\text{eve}$ and $\overline{N}_\text{eve}$ are larger than 100 for each bin.
Here we assume that the errors of the probabilities are given by
\begin{equation}
	\sigma_\text{err}^{(P)} = \frac{\sqrt{N_\text{eve}}}{N_\text{tot}}, \qquad \sigma_\text{err}^{(\overline{P})} = \frac{\sqrt{\overline{N}_\text{eve}}}{\overline{N}_\text{tot}}.
\end{equation}
Considering the propagation of error, we also assume that the error of the ratio $P / \overline{P}$ is given by
\begin{eqnarray}
	\sigma_\text{err}^{(P/\overline{P})} &=& \frac{P}{\overline{P}} \sqrt{\left(\frac{\sigma_\text{err}^{(P)}}{P}
		\right)^2 + \left(\frac{\sigma_\text{err}^{(\overline{P})}}{\overline{P}} \right)^2} \nonumber \\
	\label{eq:err_ratio}
	&=& \frac{P}{\overline{P}} \sqrt{ \frac{1}{N_\text{eve}} + \frac{1}{\overline{N}_\text{eve}}},
\end{eqnarray}
The error bars in Fig. \ref{fig:fluctuation_theorem}(a) are calculated from Eq. (\ref{eq:err_ratio}), and the solid line represents the theoretical prediction (\ref{eq:fluctuation_theorem}) which is good agreement with our simulation results.
This is the first numerical verification of the extended fluctuation theorem for inelastic collisions.

Figure \ref{fig:fluctuation_theorem}(b) exhibits the probability distributions $P$ and $\overline{P}$, in which the initial speeds are $0.009 c^\text{(t)}$ and $0.0090054 c^\text{(t)}$, respectively.
These distributions can be fitted by the Gaussians.
It is reasonable that the extended fluctuation theorem proposed by Tasaki~\cite{tasaki2006}, because our model  described by Eqs. (\ref{eq:eom_com}) and (\ref{eq:eom_vib}) assumes the separation between the translational mode and the other internal modes. 
This separation may not be satisfied in collisions based on the molecular dynamics simulation~\cite{kuninaka2009_2}.

\subsection{\label{ssec:thermal_heating} Heating during collisions}

\begin{figure}
	\includegraphics[width = 86mm]{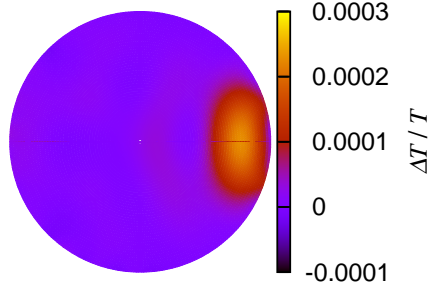}
	\caption{The heating distribution on the cross section at the instant of the impact for $v_\text{CM} = 0.1$. It should be noted that the heat distribution is plotted at the position of an undeformed sphere.}
	\label{fig:heating}
\end{figure}

Our basic equation (\ref{eq:eom_displacement}) assumes that the colliding spheres are in an isothermal state, where the collisional heating can be ignored.
To verify its validity, we estimate the amount of heating up during collisions.
It is known that the heating in linear elasticity $\Delta T(t; \bm{x})$ is proportional to the initial temperature of elastic spheres $T$ and the trace of strain tensor $\bm{\nabla} \cdot \bm{u}(t; \bm{x})$ \cite{kelvin1853, kelvin1878}
\begin{equation}
	\Delta T(t; \bm{x}) = - T \frac{3 K_\text{ad} \alpha}{c_P} \bm{\nabla} \cdot \bm{u}(t; \bm{x}),
\end{equation}
where $\alpha$ and $c_P$ are the coefficient of linear expansion and the specific heat capacity at constant pressure, and the adiabatic modulus $K_\text{ad}$ is related to the bulk modulus $K$ as
\begin{equation}
	\frac{1}{K_\text{ad}} = \frac{1}{K} - \frac{9 T \alpha^2}{c_P}.
\end{equation}
Figure \ref{fig:heating} shows the heating distribution on the cross section at the instant of the impact with the large impact speed $v_\text{CM} = 0.1$ within our framework for isothermal calculation, where we use parameters for copper $\alpha = 16.5 \times 10^{-6}$/K and $c_P = 24.5$ J/mol K, and $T = 2.14 \times 10^{-8} M (c^\text{(t)})^2$ (300 K in the physical unit).
Although we observe a little heating with the order $\Delta T/T <10^{-4}$, this small increment of the temperature is negligible, which is consistent with the isothermal assumption.
Therefore, we believe that our calculation can be used even for relatively high speed impacts.

\section{\label{sec:discussion} DISCUSSION}

Now, let us discuss our results. 
In the first part of this section (Sec. \ref{ssec:discussion_perturbation}) we develop the perturbation theory to explain the rebound processes. In the second part (Sec. \ref{ssec:discussion_mode-transfer}) we discuss the mode transfer starting from one mode excitation state.
In the last part (Sec. \ref{ssec:discussion_perspective}) we discuss future problems and perspectives.

\subsection{\label{ssec:discussion_perturbation} Perturbation theory}

In this section, we examine the perturbation theory of this system to understand the sinusoidal structure of the restitution coefficient against the initial phase (see Figs. \ref{fig:super_mode}(a) and \ref{fig:mode-transfer_v0.1}(a)).
Here we restrict our interest to the perfectly elastic case $\gamma = 0$ at $T = 0$, though we can easily extend our theory to the dissipative case i.e. $\gamma>0$ and finite T.

For perfectly elastic case, the energy conservation law leads to the simple relation for the restitution coefficient
\begin{equation} \label{eq:energy-conservation}
	e^2 = 1 - \sum_{i, n \ell m} \frac{\Delta H_{i, n \ell m} } {H_\text{CM}(0)}.
\end{equation}
Thus, if we know the excitation $\Delta H_{i, n \ell m}$, we can determine the restitution coefficient $e$.
% from Eqs. (\ref{eq:total-vib-energy}) and (\ref{eq:energy-conservation}).

First, we assume that the time evolutions of the center of mass $z_\text{CM}(t)$ and the vibrational mode $Q_{n \ell m}(t)$ are scaled by $t_\text{CM} \equiv R_\text{eff} / v_\text{CM}(0)$ and $t_\text{vib} \equiv R_\text{eff} / c^\text{(t)}$, respectively.
We introduce dimensionless variables using these time units, and the reduced radius $R_\text{eff}$ and mass $M_\text{eff}$.
Then, equations of motion (\ref{eq:eom_com}) and (\ref{eq:eom_vib}) are rewritten as
\begin{eqnarray}\label{eq:Q_tilde}
	\frac{\mathrm{d}^2 \tilde{z}_\text{CM}}{\mathrm{d} \tilde{t}_\text{CM}^2}
	+ \frac{\partial \tilde{V}(\tilde{z}_\text{CM}, \{\tilde{Q}_{i', n' \ell' m'}\})} {\partial \tilde{z}_\text{CM}} &=& 0, \\
	\frac{\mathrm{d}^2 \tilde{Q}_{i, n \ell m}}{\mathrm{d} \tilde{t}_\text{vib}^2} + \tilde{\omega}_{i, n \ell}^2 \tilde{Q}_{i, n \ell m}
	&=& - \varepsilon^2 \frac{1}{\tilde{M}_i} \frac{\partial \tilde{V}(\tilde{z}_\text{CM}, \{\tilde{Q}_{i', n' \ell' m'}\})} {\partial \tilde{Q}_{i, n \ell m}},
\end{eqnarray}
where $\tilde{z}_\text{CM} \equiv z_\text{CM} / R_\text{eff}$, $\tilde{Q}_{i, n \ell m} \equiv Q_{n \ell m} / R_\text{eff}$, $\tilde{t}_\text{CM} \equiv t / t_\text{CM}$, $\tilde{t}_\text{vib} \equiv t / t_\text{vib}$, $\tilde{\omega}_{i, n \ell} \equiv \omega_{n \ell} t_\text{vib}$, $\tilde{M}_i \equiv M_i / M_\text{eff}$ and $\tilde{V}[\tilde{z}_\text{CM}, \{\tilde{Q}_{i, n \ell m}\}] \equiv V[z_\text{CM}(t), \{Q_{n \ell m}(t)\}] / M_\text{eff} v_\text{CM}(0)^2$ are dimensionless variables.
Here we introduce the expansion parameter $\varepsilon \equiv v_\text{CM}(0) / c^\text{(t)}$, and expand $\tilde{z}_\text{CM}$ and $\tilde{Q}_{i, n \ell m}$ as:
\begin{eqnarray}
	\tilde{Q}_{i, n \ell m} &=& \tilde{Q}_{i, n \ell m}^{(0)} + \varepsilon \tilde{Q}_{i, n \ell m}^{(1)}
	+ \varepsilon^2 \tilde{Q}_{i, n \ell m}^{(2)} + \dotsb, \\
	\tilde{z}_\text{CM} &=& \tilde{z}_\text{CM}^{(0)} + \varepsilon \tilde{z}_\text{CM}^{(1)}
	+ \varepsilon^2 \tilde{z}_\text{CM}^{(2)} + \dotsb,
\end{eqnarray}
where $\tilde{z}_\text{CM}^{(j)}$ and $\tilde{Q}_{i, n \ell m}^{(j)}$ are $j$th order expansion coefficients.
We adopt that these coefficients are initially zero except for unperturbed coefficients, i.e. $\tilde{z}_\text{CM}(0) = \tilde{z}_\text{CM}^{(0)}(0)$ and $\tilde{Q}_{i, n \ell m}(0) = \tilde{Q}_{i, n \ell m}^{(0)}(0)$.
Then, the unperturbed equations are given by
\begin{eqnarray} \label{eq:0th_com}
	\frac{\mathrm{d}^2 \tilde{z}_\text{CM}^{(0)}}{\mathrm{d} \tilde{t}_\text{CM}^2}
	+ \frac{\partial \tilde{V}(\tilde{z}_\text{CM}^{(0)}, 0)}{\partial \tilde{z}_\text{CM}} &=& 0, \\
	\label{eq:0th_vib}
	\frac{\mathrm{d}^2 \tilde{Q}_{i, n \ell m}^{(0)}}{\mathrm{d} \tilde{t}_\text{vib}^2}
	+ \tilde{\omega}_{i, n \ell}^2 \tilde{Q}_{i, n \ell m}^{(0)} &=& 0,
\end{eqnarray}
where we have assumed that $\{\tilde{Q}_{i, n \ell m}^{(0)}\}$ is negligible in the potential $V$ to be consistent with the linear theory of elasticity.
The solution of Eq. (\ref{eq:0th_com}) is immediately given by
\begin{equation}
	\tilde{t}_\text{CM} = 
	\begin{cases}
		\displaystyle \int_{\tilde{z}_\text{CM}^{(0)}(0)}^{\tilde{z}_\text{CM}^{(0)}} \frac{\mathrm{d}x}{\sqrt{1 - 2 \tilde{V}(x, 0)}}
		& (\tilde{t}_\text{CM} \leq \tilde{t}_\text{CM}^\text{col}) \\
		\displaystyle \int_{\tilde{z}_\text{CM}^{(0)}(0)}^{\tilde{z}_\text{CM}^\text{col}} \frac{\mathrm{d}x}{\sqrt{1 - 2 \tilde{V}(x, 0)}}
		- \displaystyle \int_{\tilde{z}_\text{CM}^\text{col}}^{\tilde{z}_\text{CM}^{(0)}} \frac{\mathrm{d}x}{\sqrt{1 - 2 \tilde{V}(x, 0)}}
		& (\tilde{t}_\text{CM} > \tilde{t}_\text{CM}^\text{col})
	\end{cases},
\end{equation}
where $\tilde{z}_\text{CM}^\text{col} \equiv \tilde{z}_\text{CM}^{(0)}(\tilde{t}_\text{CM}^\text{col})$ are determined by the condition $1 - 2 \tilde{V}(\tilde{z}_\text{CM}^\text{col}, 0) = 0$.
From Eq. (\ref{eq:0th_vib}) $\tilde{Q}_{i, n \ell m}^{(0)}$ is just a solution of the equation for a harmonic oscillator.
The first order $\tilde{Q}_{i, n \ell m}^{(1)}$ is always zero because $\tilde{Q}^{(1)}_{i,nlm}$ satisfies the equation of the harmonic oscillator under the initial condition we introduced.
The second order equation for the internal vibration is 
\begin{equation}\label{eq:second_vib}
	\frac{\mathrm{d}^2 \tilde{Q}_{i, n \ell m}^{(2)}}{\mathrm{d} \tilde{t}_\text{vib}^2}
	+ \tilde{\omega}_{i, n \ell}^2 \tilde{Q}_{i, n \ell m}^{(2)}
	= - \frac{1}{\tilde{M}_i} \frac{\partial \tilde{V}(\tilde{z}_\text{CM}^{(0)}, 0)}{\partial \tilde{Q}_{i, n \ell m}},
\end{equation}
where we also ignore $\{\tilde{Q}_{i, n \ell m}^{(0)}\}$ in the potential $V$.
The Solution $\tilde{Q}_{i, n \ell m}^{(2)}$ of Eq. (\ref{eq:second_vib}) is given by
\begin{equation} \label{eq:second_vib}
	\tilde{Q}_{i, n \ell m}^{(2)}(\tilde{t}_\text{vib})
	= - \frac{1}{\tilde{M}_i \tilde{\omega}_{i, n \ell}} \int_0^{\tilde{t}_\text{vib}} \mathrm{d}t'
	\frac{\partial \tilde{V}(\tilde{z}_\text{CM}^{(0)}(t'), 0)}{\partial \tilde{Q}_{i, n \ell m}}
	\sin \tilde{\omega}_{i, n \ell} (\tilde{t}_\text{vib} - t').
\end{equation}
Therefore, the vibrational energy coefficients $\tilde{H}_{i, n \ell m}^{(2)} = \dot{\tilde{Q}}_{i, n \ell m}^{(0)} \dot{\tilde{Q}}_{i, n \ell m}^{(2)} + \tilde{\omega}_{i, n \ell}^2 \tilde{Q}_{i, n \ell m}^{(0)} \tilde{Q}_{i, n \ell m}^{(2)}$ and $\tilde{H}_{i, n \ell m}^{(4)} = (\dot{\tilde{Q}}_{i, n \ell m}^{(2)})^2 / 2 + (\tilde{\omega}_{i, n \ell} \tilde{Q}_{i, n \ell m}^{(2)})^2 / 2$ are, respectively, reduced to
\begin{eqnarray} \label{eq:second_energy}
	\tilde{H}_{i, n \ell m}^{(2)}(\tilde{t}_\text{vib}) &=& - 2 \sqrt{\tilde{H}_{i, n \ell m}^{(0)}(0) \tilde{H}_{i, n \ell m}^{(4)}(\tilde{t}_\text{vib})}
		\cos (\alpha_{i, n \ell m}(0) + \tilde{\omega}_{i, n \ell} \tilde{t}_\text{vib} - \beta_{i, n \ell m}(\tilde{t}_\text{vib})) \\
	\tilde{H}_{i, n \ell m}^{(4)}(\tilde{t}_\text{vib}) &=& \frac{1}{2 \tilde{M}_i^2} \left| \int_0^{\tilde{t}_\text{vib}} \mathrm{d}t'
	\frac{\partial \tilde{V}(\tilde{z}_\text{CM}^{(0)}(t'), 0)}{\partial \tilde{Q}_{i, n \ell m}} e^{i \tilde{\omega}_{i, n \ell} t'} \right|^2,
\end{eqnarray}
where $\beta_{i, n \ell m}(\tilde{t}_\text{vib})$ is determined by (see Appendix \ref{sec:app_perturbation-calc})
\begin{equation}
	\sin \beta_{i, n \ell m}(\tilde{t}_\text{vib}) = - \frac{\tilde{\omega}_{i, n \ell} \tilde{Q}_{i, n \ell m}^{(2)}(\tilde{t}_\text{vib})}
		{\sqrt{2 \tilde{H}_{i, n \ell m}^{(4)}(\tilde{t}_\text{vib})}}, \qquad
	\cos \beta_{i, n \ell m}(\tilde{t}_\text{vib}) = - \frac{\dot{\tilde{Q}}_{i, n \ell m}^{(2)}(\tilde{t}_\text{vib})}
		{\sqrt{2 \tilde{H}_{i, n \ell m}^{(4)}(\tilde{t}_\text{vib})}}.
\end{equation}
From Eqs. (\ref{eq:energy-conservation}), (\ref{eq:second_energy}) and (\ref{eq:perturb_beta}), we obtain the sinusoidal behavior of the restitution coefficient against the initial phase $\alpha_{i, n \ell m}(0)$
\begin{eqnarray}
	e^2 &=& 1 + 4 \sum_{i, n \ell m} \sqrt{\tilde{H}_{i, n \ell m}^{(0)}(0) \tilde{H}_{i, n \ell m}^{(4)}(t_f)} \cos
		\left( \alpha_{i, n \ell m}(0) + \frac{\tilde{\omega}_{i, n \ell} t_f}{2} \right) + O(\varepsilon^2) \nonumber \\
	\label{eq:perturb_sinusoid}
	&=& 1 + 2 \sqrt{2} \sum_{i, n \ell m} \sqrt{\tilde{H}_{i, n \ell m}^{(4)}(t_f)} \dot{\tilde{Q}}_{i, n \ell m}^{(0)}(t_f / 2) + O(\varepsilon^2),
\end{eqnarray}
where $t_f$ is the duration of the interaction.
Equation (\ref{eq:perturb_sinusoid}) implies that the restitution coefficient can exceed unity if $\dot{\tilde{Q}}_{i, n \ell m}^{(0)}(t_f / 2) > 0$ or the sphere expands to the axial direction at the instant $t_f / 2$, where the amplitude is proportional to the square root of the excitation energy $\sqrt{\tilde{H}_{i, n \ell m}^{(4)}(t_f)}$.

\begin{figure}
	\begin{minipage}{0.5\columnwidth}\hspace{-\columnwidth}(a)
		\begin{center}
			\includegraphics[clip, width=\columnwidth]{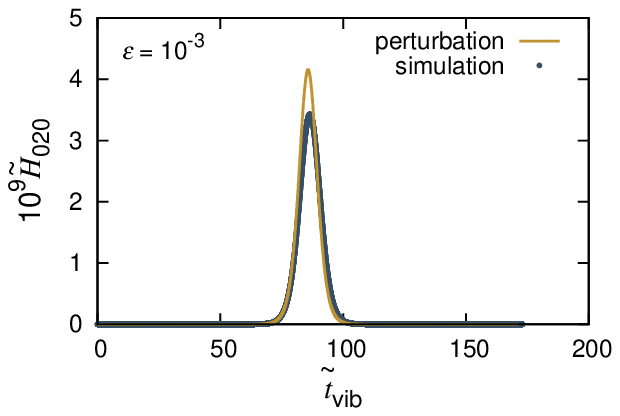}
		\end{center}
	\end{minipage}%
	\begin{minipage}{0.5\columnwidth}\hspace{-\columnwidth}(b)
		\begin{center}
			\includegraphics[clip, width=\columnwidth]{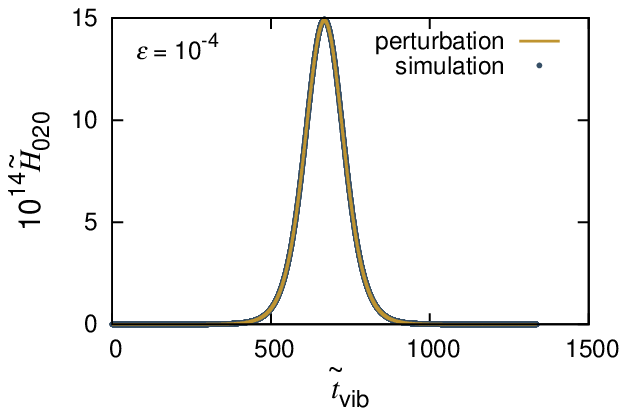}
		\end{center}
	\end{minipage}
	\caption{The time evolutions of the quadrupole mode energy $\tilde{H}_{n \ell m}$ for (a) $\epsilon = 10^{-3}$ and (b) $\epsilon = 10^{-4}$, respectively.}
	\label{fig:perturbation}
\end{figure}

Finally we compare this perturbation theory with our simulation to verify the validity of the theory.
First we numerically solve Eq. (\ref{eq:0th_com}), and then use Eq. (\ref{eq:second_vib}) to obtain the perturbative solution.
%Here we ignore the solid viscosity and the initial excitation, i.e. $\gamma = 0$ and $T = 0$.
Figures \ref{fig:perturbation} exhibits the time evolution of the quadrupole mode energy $\tilde{H}_{n \ell m}$ for (a) $\epsilon = 10^{-3}$ and (b) $\epsilon = 10^{-4}$, where we restrict our interest to the case that the interaction is only characterized by repulsion force.
We find that these are in good agreement with each other.
The agreements are also found in the other eigenmodes.

Note that our perturbation results fail to reproduce our simulation results even for small $\epsilon$ if there exists the attractive interaction, because the existence of sticking force affects the unperturbative solution.

\subsection{\label{ssec:discussion_mode-transfer} Mode transfer induced by collisions}

\begin{figure}
	\begin{minipage}{0.5\columnwidth}\hspace{-\columnwidth}(a)
		\begin{center}
			\includegraphics[clip, width=\columnwidth]{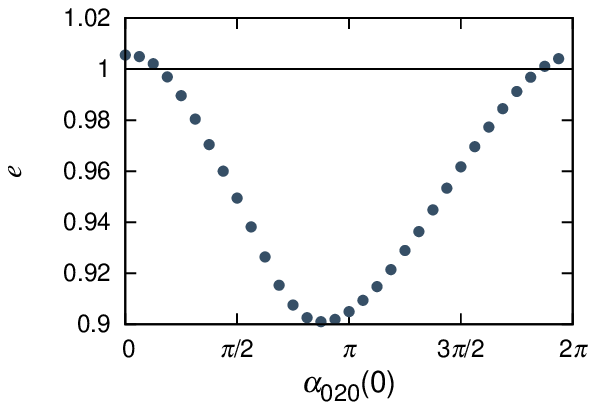}
		\end{center}
	\end{minipage}%
	\begin{minipage}{0.5\columnwidth}\hspace{-\columnwidth}(b)
		\begin{center}
			\includegraphics[clip, width=\columnwidth]{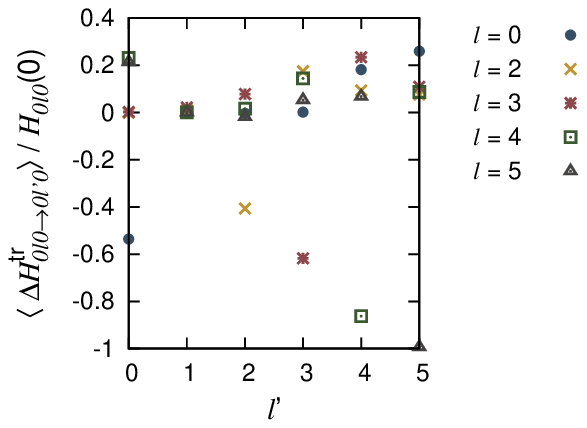}
		\end{center}
	\end{minipage}
	\caption{(a) The restitution coefficient as a function of the initial phase of the quadrupole mode and (b) phase averaged mode transfer $\langle \Delta H^\text{tr}_{0 \ell 0 \to 0 \ell' 0} \rangle / H_{0 \ell 0}(0)$ with fixing $v_\text{CM}(0) = 0.1 c^\text{(t)}$ and $H_{0 \ell 0}(0) = 0.05 H_\text{CM}(0)$.}
	\label{fig:mode-transfer_v0.1}
\end{figure}

In this subsection we study the mechanism of mode transfer during collisions.
We numerically solve Eqs. (\ref{eq:eom_com}) and (\ref{eq:eom_vib}) under only one mode excited before the collision, and we calculate the subtracted mode transfer $\Delta H^\text{tr}_{n \ell m \to n' \ell' m'}$:
\begin{equation}
	\Delta H^\text{tr}_{n \ell m \to n' \ell' m'} \equiv \Delta H_{n \ell m \to n' \ell' m'} - \Delta H_{n' \ell' m'},
\end{equation}
where mode numbers w/o primes represent the final and the initial excited modes, respectively.
Here, $\Delta H_{n' \ell' m'}$ and $\Delta H_{n \ell m \to n' \ell' m'}$ are, respectively, the energy transfer at $T=0$ and the energy transfer with the initial excitation of $(n, \ell, m)$ mode.
Note that the restitution coefficient depends on the initial phase as well as the initial excitation mode.
Figure \ref{fig:mode-transfer_v0.1}(a) shows the restitution coefficient against the initial phase of the quadrupole mode for the initial speed $v_\text{CM}(0) = 0.1 c^\text{(t)}$ and the initial excitation energy of this mode $H_{020}(0) = 0.05 H_\text{CM}(0)$.
The super rebound processes for $e>1$ can be found for small $\alpha_{020}(0)$.
Here we investigate the phase averaged mode transfer to avoid the initial phase dependence:
\begin{equation}
	\langle \Delta H^\text{tr}_{n \ell m \to n' \ell' m'} \rangle_{\alpha_{n \ell m}} \equiv \frac{1}{2 \pi} \int_0^{2 \pi} \mathrm{d}\alpha_{n \ell m} \Delta H^\text{tr}_{n \ell m \to n' \ell' m'}.
\end{equation}
Figure \ref{fig:mode-transfer_v0.1}(b) shows $\langle \Delta H^\text{tr}_{0 \ell 0 \to 0 \ell' 0} \rangle / H_{0 \ell 0}(0)$ for $v_\text{CM}(0) = 0.1 c^\text{(t)}$ and $H_{0 \ell 0}(0) = 0.05 H_\text{CM}(0)$.
The large negative value in diagonal elements means that the initial excitation energy is transferred into the other modes.
We also find that the off-diagonal elements just nearby the diagonal elements are larger than the other off-diagonal elements, which suggest that the excitation energy transfer between the nearest neighbor mode.
We find that the breathing mode ($\ell = 0$) is strongly coupled with 16-pole ($\ell = 4$) and 32-pole ($\ell = 5$) which eigenfrequencies are nearly equal to that of the breathing mode.
The dipole mode ($\ell = 1$) is decoupled with any other modes (the diagonal element is positive or nearly equal to zero), though its eigenfrequency is nearly equal to that of the octopole mode.
These results are also observed for $v_\text{CM}(0) = 0.01 c^\text{(t)}$.

%\subsection{\label{ssec:discussion_} Fluctuations and Relaxation Time}

%Although the displacement $\bm{u}_i (i = 1, 2)$ must be understood as the sums of the values of the values of the corresponding quantities in the main motion of the elastic sphere and its fluctuations~\cite{landau1959_fluid}, we ignore 
%the fluctuations because the thermal relaxation time is sufficiently longer than the contact duration in our study (see Appendix \ref{sec:app_fdr}).

\subsection{\label{ssec:discussion_perspective} Future perspectives}

In this subsection, we briefly summarize the future perspectives of our study.
Although we restrict our interest to the case of normal head on collisions of visco-elastic spheres in this paper, there are various interesting phenomena for oblique collisions.
For example, Saitoh \textit{et al}. performed the molecular dynamics simulation of the oblique collision between nanoclusters and found that the restitution coefficient of becomes negative for large incident angles~\cite{saitoh2010}.
In this anomalous collision, it is essential that the duration of contact is finite for nanocluster collisions.
% and the contact surface rotates during the contact.
%The negative restitution coefficient may be observed in our model, in which the restitution coefficient may behave more complexly because of the eigenmodes of the elastic sphere as discussed in Sec. \ref{ssec:athermal_perfect}.
To describe oblique collisions in terms of our model, we need to add the torsional modes, though the friction coefficient is necessary in the extension of the macroscopic model.

Although we only adopt the linear theory of elasticity for colliding spheres, nonlinear effects including plastic deformation and fragmentation also play crucial roles to understand the physics of collisions.
In particular, the structural phase transition caused by high speed collisions would be important to understand the physical mechanism of plastic deformation and fragmentation~\cite{johnson1985, valentini2007}.
Needless to say, the nonlinearity becomes dominant for fast collisions in which the impact speed is comparable to the sound speed of colliding bodies.

\section{\label{sec:conclusion} CONCLUSION}
In this paper, we have performed the simulation of head-on collisions based on an isothermal visco-elastic model.
%We have investigated both the athermal ($T = 0$) collision and the collision with thermally activated spheres.
%For the athermal case, 
We have investigated the restitution coefficient against the impact speed, ranging from slow $0.001 c^\text{(t)}$ to $0.4 c^\text{(t)}$, and found the oscillatory behavior in their relationship if the solid viscosity is sufficiently small.
We have confirmed that the oscillation arises from the combination of the contact duration and the eigen-frequencies of the elastic sphere.
This oscillation disappears as the solid viscosity is strong.
%It is expected that the quadrupole mode which represents the ellipsoidal deformation and has the lowest eigen-frequency is the most excited for the 
%Thus, what mode is most excited by the collision changes corresponding to the duration, or the impact speed.

We have also investigated collisions between a thermally activated elastic sphere and a flat wall.
When the impact speed of the colliding sphere is nearly equal to or slower than the thermal speed,
we have confirmed the existence of super rebounds if the attraction is reduced.
We have confirmed the existence of the fluctuation theorem for collisions of thermal activated spheres.
We have also found the sinusoidal structure of the restitution coefficient as a function of the initial phase of the eigenmodes.
%of the 16-pole mode with fixing the impact speed, in which the 16-pole mode is most excited.
This oscillation can be understood by the perturbation theory if there is no attractive force between the sphere and the wall.
\if0
We have confirmed that the ratio of the super rebound to the inelastic collision probabilities in our simulation is in good agreement with the theoretical prediction of the fluctuation theorem of inelastic collisions.
This is the first numerical verification of the fluctuation theorem of inelastic collisions.
\fi

\begin{acknowledgments}
We thank M. Sano, H. Kuninaka, K. Saitoh, and F. Kun for helpful discussion, and T. W. Kranz for his introduction of Ref. \cite{aspelmeier2000}.
This work is partially supported by the Grant-in-Aid of MEXT (Grants No. 25287098).
\end{acknowledgments}

\appendix
\section{\label{sec:app_wave-equation} THE VISCO-ELASTIC WAVE EQUATION}

In this Appendix, we derive the visco-elastic wave equation of isothermal spheres (\ref{eq:eom_displacement}) without the external potential.
First, let us consider the free energy density within the framework of the linear theory of elasticity~\cite{landau1959}
\begin{equation}\label{eq:free}
f(T, u) = f_0(T) - K (T - T_0) \alpha_{ij} u_{ij} + \frac{1}{2} \lambda_{ijkl} u_{ij} u_{kl},
\end{equation}
where $K$ is the bulk modulus, $\alpha_{ij}$ is the coefficient of thermal expansion, $\lambda_{ijkl}$ is the elastic modulus tensor and $u_{ij}$ is the strain tensor
\begin{equation}\label{eq:strain}
u_{ij} = \frac{1}{2} \left( \partial_i u_j + \partial_j u_i \right).
\end{equation}
The first term on the right hand side of Eq. (\ref{eq:free}) is independent of elastic deformation $u_{ij}$.
The second term on the right hand side represents free thermal expansion of the sphere from a base state at the temperature $T_0$.
We ignore this effect in our simulation because the heat up caused by a collision is small (see Sec. \ref{ssec:thermal_heating}).
For isotropic spheres, the elastic modulus tensor is given by
\begin{equation}
\lambda_{ijkl} = \lambda \delta_{ij} \delta_{kl} + \mu \left( \delta_{ik} \delta_{jl} + \delta_{il} \delta_{jk} \right),
\end{equation}
where $\lambda$ and $\mu$ are Lam\'e coefficients.

Thus, the free energy of isothermal elastic spheres is reduced to
\begin{equation}\label{eq:free2}
	f(T, u) = f_0(T) + \left( \frac{1}{2} \lambda u_{ii} u_{jj} + \mu u_{ij} u_{ij} \right).
\end{equation}
Here, the stress tensor $\sigma_{ij}^\text{el}$ is given by
%We can easily derive Eq. (\ref{eq:wave-equation}) using this free energy, Eq. (\ref{eq:strain}) and the relations
\begin{equation} \label{eq:stress_el}
	\sigma_{ij}^\text{el} = \left( \frac{ \partial f(T, u) }{ \partial u_{ij} } \right)_T = \lambda \delta_{ij} u_{kk} + 2 \mu u_{ij}.
%	\nonumber \\
%	\label{eq:stress_el}
%	&=& \lambda \delta_{ij} \partial_k u_k + \mu (\partial_i u_{j} + \partial_j u_{i}).
\end{equation}
Here we also consider the dissipative stress tensor $\sigma_{ij}^\text{dis}$ for isotropic bodies \cite{landau1959}
\begin{equation} \label{eq:stress_dis}
	\sigma_{ij}^\text{dis} = \lambda' \frac{\partial}{\partial t} \delta_{ij} \partial_k u_k + \mu' \frac{\partial}{\partial t} (\partial_i u_{j} + \partial_j u_{i}),
\end{equation}
where $\lambda'$ and $\mu'$ are the solid viscosity coefficients.
Then the equation of the deformation is written as
\begin{eqnarray}
	\rho \frac{\partial^2 u_i}{\partial t^2} &=& \partial_j (\sigma_{ij}^\text{el} + \sigma_{ij}^\text{dis}) \nonumber \\
	&=& \left\{ \lambda + \mu + (\lambda' + \mu') \frac{\partial}{\partial t} \right\} \partial_i \partial_j u_{j}
		+ \left( \mu + \mu' \frac{\partial}{\partial t} \right) \partial_j^2 u_{i},
\end{eqnarray}
or 
\begin{eqnarray}
	\frac{\partial^2 \bm{u}}{\partial t^2} &=& \left( \frac{\lambda + \mu}{\rho} + \frac{\lambda' + \mu'}{\rho}
		\frac{\partial}{\partial t} \right) \bm{\nabla} \bm{\nabla} \cdot \bm{u} + \left( \frac{\mu}{\rho} +
		\frac{\mu'}{\rho} \frac{\partial}{\partial t} \right) \bm{\nabla}^2 \bm{u} \nonumber \\
%	&=& \left( \frac{\lambda + 2 \mu}{\rho} + \frac{\lambda' + 2 \mu'}{\rho}
%		\frac{\partial}{\partial t} \right) \bm{\nabla} \bm{\nabla} \cdot \bm{u} - \left( \frac{\mu}{\rho} +
%		\frac{\mu'}{\rho} \frac{\partial}{\partial t} \right) \bm{\nabla} \times (\bm{\nabla} \times \bm{u}) \nonumber \\
	\label{eq:wave-equation}
	&=& \left(c^{(\ell)} \right)^2 \left( 1 + \gamma^{(\ell)} \frac{\partial}{\partial t} \right)
		\bm{\nabla} \bm{\nabla} \cdot \bm{u} - \left(c^\text{(t)} \right)^2 \left( 1 +
		\gamma^\text{(t)} \frac{\partial}{\partial t} \right) \bm{\nabla} \times (\bm{\nabla} \times \bm{u}).
\end{eqnarray}
%I have used the identity $\bm{\nabla}^2 \bm{u} = \bm{\nabla} \bm{\nabla} \cdot \bm{u} - \bm{\nabla} \times (\bm{\nabla} \times \bm{u})$ in the second line.
%In the final line, 
We have introduced
\begin{eqnarray}
	c^{(\ell)} &=& \sqrt{\frac{\lambda + 2 \mu}{\rho}}, \\
	c^\text{(t)} &=& \sqrt{\frac{\mu}{\rho}},
\end{eqnarray}
and 
\begin{eqnarray} \label{eq:viscosity_lon}
	\gamma^{(\ell)} &\equiv& \frac{\lambda' + 2 \mu'}{\lambda + 2 \mu}, \\
	\label{eq:viscosity_tra}
	\gamma^\text{(t)} &\equiv& \frac{\mu'}{\mu}.
\end{eqnarray}
%where $\sigma_{ij}$ is ij component of the stress tensor.

%\input{../620app_stressfree-solution.tex}
\section{\label{sec:app_stressfree-solution} THE STRESS-FREE SOLUTIONS OF VISCO-ELASTIC SPHERES}

%Before considering collisions, we solve the wave equation of viscoelastic spheres under stress-free conditions,  which is given by (see Appendix \ref{sec:app_wave-equation})
%\begin{eqnarray}
%	\ddot{\bm{u}}(t; \bm{x}) &=& \left(c^{(\ell)} \right)^2
%		\nabla \bm{\nabla} \cdot \left\{\bm{u}(t; \bm{x}) + \gamma^{(\ell)} \dot{\bm{u}}(t; \bm{x}) \right\} \nonumber \\
%	&& - \left(c^\text{(t)} \right)^2 \bm{\nabla} \times
%		\left[\nabla \times \left\{\bm{u}(t; \bm{x}) + \gamma^\text{(t)} \dot{\bm{u}}(t; \bm{x}) \right\} \right],
%\end{eqnarray}
%where $\gamma^{(\ell)}$ and $\gamma^\text{(t)}$ are the longitudinal and horizontal viscosity coefficients, respectively.
In this appendix, we solve the wave equation (\ref{eq:wave-equation}) which is equivalent to Eq. (\ref{eq:eom_displacement}) of visco-elastic spheres under stress-free conditions.
We now look for a special solution of the form
\begin{equation}\label{eq:sep.var.}
\bm{u}(t, \bm{x})  = e^{st} \tilde{\bm{u}}(\bm{x}),
\end{equation}
where $s$ is a complex number, corresponding to the Laplace transform without the effect of the initial condition.
Substituting Eq. (\ref{eq:sep.var.}) into Eq. (\ref{eq:wave-equation}), we obtain
\begin{equation}\label{eq:wave2}
s^2 \tilde{\bm{u}} = \left( c^{(\ell)} \right)^{2} \left( 1 + \gamma^{(\ell)} s \right) \bm{\nabla} \bm{\nabla} \cdot \tilde{\bm{u}} - \left( c^{\text{(t)}} \right)^{2} \left( 1 + \gamma^\text{(t)} s \right) \bm{\nabla} \times \left( \bm{\nabla} \times \tilde{\bm{u}} \right).
\end{equation}
To solve Eq. (\ref{eq:wave2}), we adopt the Helmholtz decomposition
\begin{equation}
\tilde{\bm{u}} = \tilde{\bm{u}}^{(\ell)} + \tilde{\bm{u}}^\text{(t)},
\end{equation}
where $\tilde{\bm{u}}^{(\ell)}$ and $\tilde{\bm{u}}^\text{(t)}$ are rotation-free and divergence-free solutions, respectively:
\begin{eqnarray}
\nabla \times \tilde{\bm{u}}^{(\ell)} &=& \bm{0}, \\
\nabla \cdot \tilde{\bm{u}}^\text{(t)} &=& 0.
\end{eqnarray}
Therefore,  $\tilde{\bm{u}}^{(\ell)}$ can be represented using one scalar potential $\Phi^{(0)}$ and $\tilde{\bm{u}}^\text{(t)}$ two scalar potentials $\Phi^{(1, 2)}$ with 
\begin{eqnarray} \label{eq:solution}
\tilde{\bm{u}}^{(\ell)} &=& \bm{\nabla} \Phi^{(0)}, \\ \label{eq:solutiont}
\tilde{\bm{u}}^\text{(t)} &=& \bm{\nabla} \times \left( \bm{x} \Phi^{(1)} \right) + \bm{\nabla} \times \left( \bm{\nabla} \times \left( \bm{x} \Phi^{(2)} \right) \right).
\end{eqnarray}
Substituting Eqs. (\ref{eq:solution}) and (\ref{eq:solutiont}) into Eq. (\ref{eq:wave2}), one can easily check that these potentials $\Phi^{(i)}$ satisfies the Helmholtz equation.
Therefore, $\Phi^{(i)}$ is given by the product of the spherical Bessel function $j_{\ell} \left( k_{n \ell}^{(i)}r \right)$, where the spherical Neumann function is automatically excluded because of the singularity at the origin, and the spherical harmonics $Y_{\ell m}(\theta, \varphi)$ in a spherical coordinate system:
\begin{equation}\label{eq:harmonic-function}
\Phi^{(i)} = \Phi^{(i)}_{n \ell m} = B^{(i)}_{n \ell m} j_{\ell} \left( k_{n \ell}^{(i)}r \right) Y_{\ell m}(\theta, \varphi),
\end{equation}
where $B^{(i)}_{n \ell m}$ are the superposition coefficients, and
\begin{eqnarray}
k_{n \ell}^{(\ell)} &\equiv& k^{(0)}_{n \ell}, \\
k_{n \ell}^\text{(t)} &\equiv& k^{(1)}_{n \ell} = k^{(2)}_{n \ell}.
\end{eqnarray}
Here, $k_{n \ell}^{(\ell)}$ and $k_{n \ell}^\text{(t)}$ respectively satisfy the dispersion relations
\begin{eqnarray}\label{eq:dispersion1}
-s_{n \ell}^2 &=& \left( c^{(\ell)} k_{n \ell}^{(\ell)} \right)^{2} \left( 1 + \gamma^{(\ell)} s_{n \ell} \right), \\ \label{eq:dispersion2}
-s_{n \ell}^2 &=& \left( c^{\text{(t)}} k_{n \ell}^\text{(t)} \right)^{2} \left( 1 + \gamma^\text{(t)} s_{n \ell} \right).
\end{eqnarray}
Substituting Eq. (\ref{eq:harmonic-function}) into Eq. (\ref{eq:solution}) with the aid of the differential equation for the spherical Bessel function, we obtain
\begin{eqnarray} \label{eq:stressfree-solution}
	\tilde{\bm{u}}(\bm{x}) = \tilde{\bm{u}}_{n \ell m}(\bm{x}) &=& \left[ B^{(0)}_{n \ell m} \frac{\mathrm{d}j_{\ell} \left( k_{n \ell}^{(\ell)}r \right)}
		{\mathrm{d}r} + B^{(2)}_{n \ell m} \ell (\ell + 1) \frac{j_{\ell} \left( k_{n \ell}^{\text{(t)}} r \right)}{r} \right]
		Y_{\ell m}(\theta, \varphi) \bm{e}_{r} \nonumber \\
	&& + \left[ B^{(0)}_{n \ell m} j_{\ell} \left( k_{n \ell}^{(\ell)} r \right) + B^{(2)}_{n \ell m} \frac{\mathrm{d} \left\{
		r j_{\ell} \left(k_{n \ell}^{\text{(t)}} r \right) \right\}}{\mathrm{d}r} \right] \bm{\nabla} Y_{\ell m}(\theta, \varphi) \nonumber \\
 	&& - B_{n \ell m}^{(1)} r j_\ell \left( k_{n \ell}^\text{(t)} r \right) \bm{e}_r \times \bm{\nabla} Y_{\ell m} (\theta, \varphi).
\end{eqnarray}
Note that $\bm{e}_r$, $\nabla Y_{\ell m}$ and $\bm{e}_r \times \bm{\nabla} Y_{\ell m}$ are, respectively, orthogonal to each other.

Here it should be noted that one can reduce the dispersion relations (\ref{eq:dispersion1}) and (\ref{eq:dispersion2}) to simpler forms by rewriting the dispersion relations in real and imaginary parts, separately, if
\begin{equation}
	0 < \gamma^{(\ell)} < \frac{2}{c^{(\ell)} k_{n \ell}^{(\ell)}}, \qquad 0 < \gamma^\text{(t)} < \frac{2}{c^\text{(t)} k_{n \ell}^\text{(t)}}.
\end{equation}
The imaginary part of the dispersion relation become
\begin{equation}\label{eq:imaginary}
\left( c^{(\ell)} k_{n \ell}^{(\ell)} \right)^{2} \gamma^{(\ell)} = \left( c^{\text{(t)}} k_{n \ell}^\text{(t)} \right)^{2} \gamma^\text{(t)},
\end{equation}
and the real part of the dispersion relations with the aid of Eq. (\ref{eq:imaginary}) is reduced to
\begin{equation}\label{eq:real}
c^{(\ell)} k_{n \ell}^{(\ell)} = c^{\text{(t)}} k_{n \ell}^\text{(t)} \equiv \omega_{n \ell},
\end{equation}
where $\omega_{n \ell}$ is the eigen frequency.
From Eqs. (\ref{eq:imaginary}) and (\ref{eq:real}), we obtain a counter intuitive relation
\begin{equation} \label{eq:viscosity_identity}
\gamma^{(\ell)} = \gamma^\text{(t)} \equiv \gamma.
\end{equation}
This result is remarkable, because there is only one solid viscosity.

Now, let us consider the solution of Eq. (\ref{eq:wave2}) under the stress-free boundary condition
\begin{equation}\label{eq:boundary}
	(\bm{F}_r)_i \equiv \frac{x_j}{r} \sigma_{ij}^\text{el}(R, \theta, \varphi) = 0.
\end{equation}
With the aid of Eq. (\ref{eq:stress_el}) we can rewrite $\bm{F}_r$ as
\begin{equation} \label{eq:stress2}
	\bm{F}_r = \lambda \bm{\nabla} \cdot \bm{u} \bm{e}_r
		+ \mu \left( \bm{\nabla} u_r + \frac{u_r}{r} \bm{e}_r -\frac{\bm{u}}{r} + \frac{\partial \bm{u}}{\partial r} \right).
\end{equation}
Substituting Eq. (\ref{eq:stressfree-solution}) into Eq. (\ref{eq:stress2}), we find
\begin{eqnarray} \label{eq:r-stress}
	\frac{\bm{F}_r}{\mu} = \frac{\bm{F}_{r,n \ell m}}{\mu}
	&=& \left[ \left( k_{n \ell}^{(\ell)} \right)^2 B^{(0)}_{n \ell m} a_{n \ell} \left(k_{n \ell}^{(\ell)}r\right)
		+ \left( k_{n \ell}^{\text{(t)}} \right)^2 B^{(2)}_{n \ell m} \ell (\ell + 1)
		b_{n \ell} \left(k_{n \ell}^{\text{(t)}}r\right) \right] Y_{\ell m}(\theta, \varphi) \bm{e}_{r} \nonumber \\
	&& + \left[ \left( k_{n \ell}^{(\ell)} \right)^2 B^{(0)}_{n \ell m} b_{n \ell}(k_{n \ell}^{(\ell)}r)
		+ \left( k_{n \ell}^{\text{(t)}} \right)^2 B^{(2)}_{n \ell m} d_{n \ell}(k_{n \ell}^{\text{(t)}}r) \right]
		r \bm{\nabla} Y_{\ell m}(\theta, \varphi) \nonumber \\
	&& + \frac{1}{2} \left( k_{n \ell}^{\text{(t)}} r \right)^2 B^{(1)}_{n \ell m} b_{n \ell}(k_{n \ell}^{\text{(t)}}r)
		\bm{e}_{r} \times \bm{\nabla} Y_{\ell m}(\theta, \varphi),
\end{eqnarray}
where
\begin{eqnarray}
a_{n \ell} (x) &=& 2 \frac{ \mathrm{d}^2 j_\ell(x) } { \mathrm{d} x^2 } - \frac{\lambda}{\mu} j_\ell (x), \\
b_{n \ell} (x) &=& 2 \frac{ \mathrm{d} } { \mathrm{d} x } \left( \frac{ j_\ell(x) } { x } \right), \\
d_{n \ell} (x) &=& 2 x \frac{ \mathrm{d}^2 j_\ell(x) } { \mathrm{d} x^2 } + (\ell - 1) (\ell + 2) \frac{j_\ell(x)}{x}.
\end{eqnarray}
The boundary condition (\ref{eq:boundary}) is also reduced to a set of the following equations
\begin{eqnarray} \label{eq:sboundary}
	\bm{A}
	\left(\begin{array}{c}
		\left( k_{n \ell}^{(\ell)} \right)^2 B^{(0)}_{n \ell m} \\
		\left( k_{n \ell}^{\text{(t)}} \right)^2 B^{(2)}_{n \ell m}
	\end{array}\right) & = & \bm{0}, \\
	b_{n \ell} \left( k_{n \ell}^{\text{(t)}} R \right) B^{(1)}_{n \ell m} & = & 0.
\end{eqnarray}
where
\begin{equation}
	\bm{A} \equiv 
	\left(\begin{array}{cc}
		a_{n \ell} \left(k_{n \ell}^{(\ell)} R \right) & \ell (\ell + 1) b_{n \ell} \left(k_{n \ell}^{\text{(t)}} R \right) \\
		b_{n \ell} \left(k_{n \ell}^{(\ell)} R \right) & d_{n \ell}\left(k_{n \ell}^{\text{(t)}} R \right)
	\end{array}\right).
\end{equation}
Thus, there are two types of modes; the spheroidal modes 
\begin{eqnarray}
	\left(\begin{array}{c}	B^{(0)}_{n \ell m} \\
						B^{(2)}_{n \ell m}	\end{array}\right) & \neq & \bm{0}, \\
	\label{eq:spheroidal-condition}
	B^{(1)}_{n \ell m} & = & 0,
\end{eqnarray}
and the torsional modes
\begin{eqnarray}
\left(\begin{array}{c}
B^{(0)}_{n \ell m} \\
B^{(2)}_{n \ell m}
\end{array}\right) & = & \bm{0}, \\
B^{(1)}_{n \ell m} & \neq & 0.
\end{eqnarray}
Therefore, the solution of the spheroidal mode is given by
\begin{eqnarray} \label{eq:spheroidal-solution}
	\tilde{\bm{u}}_{n \ell m}^\text{(S)}(\bm{x}) &=& \left[ B^{(0)}_{n \ell m} \frac{\mathrm{d}j_{\ell} \left( k_{n \ell}^{(\ell)}r \right)}
		{\mathrm{d}r} + B^{(2)}_{n \ell m} \ell (\ell + 1) \frac{j_{\ell} \left( k_{n \ell}^{\text{(t)}} r \right)}{r} \right]
		Y_{\ell m}(\theta, \varphi) \bm{e}_{r} \nonumber \\
	&& + \left[ B^{(0)}_{n \ell m} j_{\ell} \left( k_{n \ell}^{(\ell)} r \right) + B^{(2)}_{n \ell m} \frac{\mathrm{d}
		\left\{ r j_{\ell} \left(k_{n \ell}^{\text{(t)}} r \right) \right\}}{\mathrm{d}r} \right] \bm{\nabla} Y_{\ell m}(\theta, \varphi).
\end{eqnarray}
The eigenfrequency $\omega_{n \ell}$ is obtained from $\det \bm{A} = 0$, and the ratio $B^{(0)}_{n \ell m} / B^{(2)}_{n \ell m}$ is determined by Eq. (\ref{eq:sboundary}).
Using the remaining freedom, we normalize $\tilde{\bm{u}}_{n \ell m}^\text{(S)}(\bm{x})$:
\begin{equation} \label{eq:normalization}
	\int_0^R \mathrm{d}r r^2 \int_0^\pi \mathrm{d} \theta \sin \theta \int_0^{2 \pi} \mathrm{d} \varphi 
		\left| \tilde{\bm{u}}_{n \ell m}^\text{(S)}(\bm{x}) \right|^2 = \frac{4 \pi} {3} R^3.
\end{equation}

\section{\label{sec:app_fdr} Fluctuations in continuum dynamics}

As mentioned in the Introduction, the solid viscosity should be associated with the random noise term to satisfy the fluctuation-dissipation relation. 
In this appendix, we briefly summarize the form of fluctuating dissipative stress tensor which represents the random noise in the stress. 
We also estimate the critical solid viscosity at which the relaxation time originated from the solid viscosity is comparable to the duration of contact.
% the duration of contact which is much shorter than the characteristic relaxation time originated from the solid viscosity.
 
Here, we assume that the dissipative stress tensor $\sigma_{ij}^\text{dis}$ is given by (\ref{eq:stress_dis}).
In the presence of fluctuations, however, there is fluctuating local stress $\delta \sigma_{ij} ^\text{dis}$
Thus the dissipative stress is replaced by
\begin{equation} \label{eq:add-fluctuation}
	\sigma_{ij}^\text{dis} \to \sigma_{ij}^\text{dis} + \delta \sigma_{ij} ^\text{dis}.
\end{equation}
As in the case of the fluctuating hydrodynamics, the fluctuating stress satisfies the relations:
\begin{equation}
	\langle \delta \sigma_{ij} ^\text{dis} (t; \bm{x}) \rangle = 0,
\end{equation}
and~\cite{landau1959_fluid}
\begin{equation} \label{eq:fdt}
	\langle \delta \sigma_{ij} ^\text{dis} (t_1; \bm{x}_1) \delta \sigma_{k \ell} ^\text{dis} (t_2; \bm{x}_2) \rangle = 2 T \{ 2 \mu' ( \delta_{ik} \delta_{j \ell} + \delta_{i \ell} \delta_{jk}) + \lambda' \delta_{ij} \delta_{k \ell} \} \delta(t_1- t_2) \delta(\bm{x}_1 - \bm{x}_2),
\end{equation}
where we denote the statistical average by $\langle \rangle$.

Although we have introduced the fluctuating local stress in Eq. (\ref{eq:add-fluctuation}), it is not important for collisions for small $\gamma$.
Here we identify the spontaneous relaxation time $\tau_{n \ell} ^\text{(r)}$ coupled with $\gamma$.
From Eqs. (\ref{eq:sep.var.}), $\tau_{n \ell} ^\text{(r)}$ is defined by
\begin{equation}
	\tau_{n \ell} ^\text{(r)} \equiv - \frac{1} {\mathrm{Re} [s_{n \ell}] },
\end{equation}
where $\mathrm{Re} [s_{n \ell}]$ represents the real part of $s_{n \ell}$.
Combination of Eqs. (\ref{eq:dispersion1}), (\ref{eq:real}) and (\ref{eq:viscosity_identity}) leads to
\begin{equation}
	\tau_{n \ell} ^\text{(r)} = \frac{2} {\omega_{n \ell}^2 \gamma}.
\end{equation}
On the other hand, the duration of contact in the quasi-static theory is given by~\cite{landau1959}
\begin{equation}
	\tau_\text{H} = 2.87 \left( \frac{M_\text{eff} ^2} {Y_\text{eff} ^2 R_\text{eff} v_\text{CM}(0)} \right)^{1/5}.
\end{equation}
It should be noted that $\tau_\text{H} (v_\text{CM}(0))$ is nearly equal to the duration of contact of our simulation, ranging from $v_\text{CM}(0) = 0.001 c^\text{(t)}$ to $v_\text{CM}(0) = 0.4 c^\text{(t)}$.
Here we introduce the critical solid viscosity $\gamma_{n \ell} ^*$ at which $\tau_{n \ell} ^\text{(r)} = \tau_\text{H}$ is satisfied.
Thus, we obtain
\begin{equation}
	\gamma_{n \ell} ^* = \frac{0.7} {\omega_{n \ell}^2} \left( \frac{Y_\text{eff} ^2 R_\text{eff} v_\text{CM}(0)} {M_\text{eff} ^2} \right)^{1/5}.
\end{equation}
For the lowest eigenfrequency $\omega_{02} \simeq 2.65 c^\text{(t)} / R_1$ and $v_\text{CM}(0) = 0.001 c^\text{(t)}$,
the corresponding critical solid viscosity is estimated as
\begin{equation}
	\gamma_{02} ^* \simeq 0.02 R_1 / c^\text{(t)},
\end{equation}
when the target is a flat wall.
Therefore, the fluctuating stress is negligible for $\gamma \ll 0.02 R_1 / c^\text{(t)}$, which is satisfied for most of the cases we have analyzed in this paper.

\section{\label{sec:app_force} SOME EXPLICIT EXPRESSIONS}

In this appendix, we briefly summarize the explicit form of the distance (\ref{eq:distance}) for the axisymmetric case, and the potential (\ref{eq:total_interaction_twospheres}) for the hard wall limit, i.e. $c_2^\text{(t)} \to \infty$, $c_2^{(\ell)} \to \infty$ and $R_2 \to \infty$.

\subsection{The distance (\ref{eq:distance}) for the axisymmetric case}

To summarize some complicated expressions, we introduce
\begin{eqnarray}
	z(z_\text{CM}, \{Q_{i', n' \ell' m'}\}; \theta_1, \varphi_1; \theta_2, \varphi_2)
	&\equiv& z_\text{CM} - R_2 \cos\theta_2 - u_{z2}(\{Q_{2, n' \ell' m'}\}; R_2, \theta_2, \varphi_2) \nonumber \\
	& & - R_1 \cos\theta_1 - u_{z1}(\{Q_{1, n' \ell' m'}\}; R_1, \theta_1, \varphi_1), \\
	u_{xy} &\equiv& u_r \cos\theta + u_\theta \sin\theta.
\end{eqnarray} 
If the initial excitation of elastic spheres is absent, the normal head-on collision is axisymmetric, in which $u_\varphi = 0$ and $u_{xy}$ and $z(z_\text{CM}, \{Q_{i', n' \ell' m'}\}; \theta_1, \varphi_1; \theta_2, \varphi_2)$ are independent of both $\varphi_1$ and $\varphi_2$.
Then, the distance (\ref{eq:distance}) and its derivative are given by 
%Eqs. (\ref{eq:distance_general}) and (\ref{eq:diff-dis1_general}) can be reduced to
\begin{eqnarray}
	\label{eq:distance_axisym}
	r^2 &=& \{ R_2 \sin\theta_2 + u_{xy2}(\{Q_{2, n' \ell' m'}\}; R_2, \theta_2) \}^2 
		+ \{ R_1 \sin\theta_1 + u_{xy1}(\{Q_{1, n' \ell' m'}\}; R_1, \theta_1) \}^2 \nonumber \\
	&& + 2 \{ R_2 \sin\theta_2 + u_{xy2}(\{Q_{2, n' \ell' m'}\}; R_2, \theta_2) \}
		\{ R_1 \sin\theta_1 + u_{xy1}(\{Q_{1, n' \ell' m'}\}; R_1, \theta_1) \} \cos ( \varphi_1 + \varphi_2 ) \nonumber \\
	&& + \{ z(z_\text{CM}, \{Q_{i', n' \ell' m'}\}; \theta_1; \theta_2 ) \}^2, \\
	\label{eq:diff-dis1_axisym}
	\frac{1}{2} \frac{\partial r^2}{\partial Q_{1,n \ell m}} &=& \tilde{u}_{xy,n \ell m}( R_1, \theta_1, \varphi_1 )
		[ \{ R_2 \sin\theta_2 + u_{xy2}(\{Q_{2, n' \ell' m'}\}; R_2, \theta_2) \} \cos ( \varphi_1 + \varphi_2 ) \nonumber \\
	&& + \{ R_1 \sin\theta_1 + u_{xy1}(\{Q_{1, n' \ell' m'}\}; R_1, \theta_1) \} ] \nonumber \\
	&& - \tilde{u}_{\varphi,n \ell m}( R_1, \theta_1, \varphi_1 )
		\{ R_2 \sin\theta_2 + u_{xy2}(\{Q_{2, n' \ell' m'}\}; R_2, \theta_2) \} \sin ( \varphi_1 + \varphi_2 ) \nonumber \\
	&& - \tilde{u}_{z,n \ell m}( R_1, \theta_1, \varphi_1 ) z(z_\text{CM}, \{Q_{i', n' \ell' m'}\}; \theta_1; \theta_2 ).
\end{eqnarray}
Equation (\ref{eq:distance_axisym}) includes only $\varphi_1 + \varphi_2$ and Eq. (\ref{eq:diff-dis1_axisym}) includes $\varphi_1$ and $\varphi_1 + \varphi_2$.
In addition, $\phi_1$ dependence only appears in the coefficients $\tilde{u}_{xy,n \ell m}$, $\tilde{u}_{\varphi,n \ell m}$ and $\tilde{u}_{z,n \ell m}$ where the integral with respect to $\phi_1$ disappears except for $m = 0$.
Therefore, both $F_{1,n \ell m}$ and $Q_{1, n \ell m}$ for $m \neq 0$ are absent during the axisymmetric collision.
For $m = 0$, $\tilde{u}_{\varphi,n \ell 0} = 0$, and $\tilde{u}_{xy,n \ell 0}$ and $\tilde{u}_{z,n \ell 0}$ is also independent of $\varphi$.
Thus, Eq. (\ref{eq:diff-dis1_axisym}) becomes
\begin{eqnarray}
	\label{eq:coef}
	\frac{1}{2} \frac{\partial r^2}{\partial Q_{1,n \ell 0}} &=& \tilde{u}_{xy,n \ell 0}( R_1, \theta_1 )
		[ \{ R_2 \sin\theta_2 + u_{xy2}(\{Q_{2, n' \ell' 0}\}; R_2, \theta_2) \} \cos ( \varphi_1 + \varphi_2 ) \nonumber \\
	&& + \{ R_1 \sin\theta_1 + u_{xy1}(\{Q_{1, n' \ell' 0}\}; R_1, \theta_1) \} ] \nonumber \\
	&& - \tilde{u}_{z,n \ell}( R_1, \theta_1 ) z(z_\text{CM}, \{Q_{i', n' \ell' 0}\}; \theta_1; \theta_2 ).
\end{eqnarray}
$\partial r^2 / \partial Q_{1, n \ell 0}$ also depends only on $\varphi_1 + \varphi_2$.
The integration of $\varphi_1 + \varphi_2$ in $F_{1,n \ell m}$ can be excluded analytically, while we avoid writing the complicated result.
See the detailed calculation in Ref \cite{murakami}

\subsection{The potential (\ref{eq:total_interaction_twospheres}) for the hard wall limit}

In the case of the limitation $c_2^\text{(t)} \to \infty$, $c_2^{(\ell)} \to \infty$ and $R_2 \to \infty$, the potential (\ref{eq:total_interaction_twospheres}) can be reduced to
\begin{eqnarray}\label{eq:total_interaction_wall}
	V(z_\text{CM}, \bm{u}_1)
	&=& 4 \epsilon \frac{R_1^2} {d_1^2 d_2^2}
		\int^{\pi/2}_0 \mathrm{d}\theta_1 \sin\theta_1 \int^{2\pi}_0 \mathrm{d}\varphi_1
		\int^\infty_{-\infty} \mathrm{d}x_2 \int^\infty_{-\infty} \mathrm{d}y_2 \nonumber \\
	&& \left[\left(\frac{\sigma}{\sqrt{(x_2 - x_1)^2 + (y_2 - y_1)^2 + (z_2 - z_1)^2}}\right)^{12} \right. \nonumber \\
	&& \left. - \left(\frac{\sigma}{\sqrt{(x_2 - x_1)^2 + (y_2 - y_1)^2 + (z_2 - z_1)^2}}\right)^6 \right], \nonumber \\
	&=& 4 \pi \epsilon \frac{R_1^2 \sigma^2} {d^4} \int^{\pi/2}_0 \mathrm{d}\theta_1 \sin\theta_1
		\int^{2\pi}_0 \mathrm{d}\varphi_1 \left[\frac{1}{5} \left(\frac{\sigma}{z_2 - z_1}\right)^{10}
		- \frac{1}{2} \left(\frac{\sigma}{z_2 - z_1}\right)^4 \right], \nonumber \\
	&=& 4 \pi \epsilon \frac{R_1^2 \sigma^2} {d^4}
		\int^{\pi/2}_0 \mathrm{d}\theta_1 \sin\theta_1 \int^{2\pi}_0 \mathrm{d}\varphi_1 \nonumber \\
	&& \left[\frac{1}{5} \left(\frac{\sigma}{z_\text{cm,w} + R_1\cos\theta_1 + u_z(R_1, \theta_1, \varphi_1)}\right)^{10} \right. \nonumber \\
	&& \left. - \frac{1}{2} \left(\frac{\sigma}{z_\text{cm,w} + R_1\cos\theta_1 + u_z(R_1, \theta_1, \varphi_1)}\right)^4 \right],
\end{eqnarray}
where $(x_1, y_1, z_1)$ and $(x_2, y_2, z_2)$ are the positions on the surface of the sphere and the wall, respectively.
We have introduced $z_\text{cm,w}$ as the distance between the center of mass position of the sphere $1$ and the wall, and used polar coordinates to obtain the last equality.
Note that the integral $\int \varphi_1$ is just reduced to $2\pi$ in Eq. (\ref{eq:total_interaction_wall}) if the initial vibration is absent but the replacement cannot be used for the initial excited case because $u_z(R_1, \theta_1, \varphi_1)$ depends on $\varphi_1$.

\if0
\begin{equation} \label{eq:symmetric-distance-between-twospheres}
	r(\theta_1, \varphi_1; \theta_2, \varphi_2) = \sqrt{s(\theta_1; \theta_2) + w(\theta_1; \theta_2) \cos(\varphi_1 + \varphi_2)}
\end{equation}
where $s(\theta_1; \theta_2)$ and $w(\theta_1; \theta_2)$ are given by Eqs. (\ref{eq:coefficient_s}) and (\ref{eq:coefficient_w}), respectively.
Substituting Eqs.(\ref{eq:lennard-jones}) and (\ref{eq:symmetric-distance-between-twospheres}) into (\ref{eq:total_interaction_twospheres}), 
\begin{eqnarray} \label{eq:symmetric-potential}
	V(z_\text{CM}, \bm{u}_1, \bm{u}_2)
		&=& \pi \epsilon \frac{R_1^2 R_2^2} {d_1^2 d_2^2} \int^{\pi/2}_0 \mathrm{d}\theta_1 \sin\theta_1
			\int^{\pi/2}_0 \mathrm{d}\theta_2 \sin\theta_2 \nonumber \\
		&& \times \left[ s(\theta_1; \theta_2) \frac{8 \{s(\theta_1; \theta_2)\}^4
			+ 40 \{s(\theta_1; \theta_2) w(\theta_1; \theta_2)\}^2 + 15 \{w(\theta_1; \theta_2)\}^4}
			{\left[\{s(\theta_1; \theta_2)\}^2-\{w(\theta_1; \theta_2)\}^2\right]^{11/2}} \right. \nonumber \\
		&& \left. - 4 \frac{2 \{s(\theta_1; \theta_2)\}^2 + \{w(\theta_1; \theta_2)\}^2}
			{\left[\{s(\theta_1; \theta_2)\}^2 - \{w(\theta_1; \theta_2)\}^2\right]^{5/2}} \right]
\end{eqnarray}
\begin{figure}
	\includegraphics[width = 86mm]{../eps_processed/coordinate/with-wall.eps}
	\caption{A schematic picture of a collision between an isothermal elastic sphere and a flat wall.}
	\label{fig:coordinate_wall}
\end{figure}
%Here, $g$ is the cohesive parameter which changes the magnitude of cohesion between atoms \cite{awasthi2007, sakiyama2004}.
\fi

\section{\label{sec:app_relation-bri} THE THIRD TERM ON THE RIGHT HAND SIDE OF EQ. (\ref{eq:bri-force})}

Here we explain that the coefficient $\gamma$ in the third term on the right hand side of Eq. (\ref{eq:bri-force}) is identical to that used by Brilliantov \textit{et al}.~\cite{brilliantov2007}
\begin{equation}
	A = \alpha^2 \beta = \left( \frac{\lambda'}{\lambda} \right)^2 \frac{3 \lambda + 2 \mu}{3 \lambda' + 2 \mu'},
\end{equation}
where the notation corresponds to Eqs. (\ref{eq:stress_el}) and (\ref{eq:stress_dis}) instead of their notation
\begin{eqnarray}
	\sigma_{ij}^\text{el} &=& E_1\left( u_{ij} - \frac{1}{3} \delta_{ij} u_{kk} \right) + E_2 \delta_{ij} u_{kk}, \\
	\sigma_{ij}^\text{dis} &=& \eta_1 \frac{\partial}{\partial t} \left( u_{ij} - \frac{1}{3} \delta_{ij} u_{kk} \right)
		+ \eta_2 \frac{\partial}{\partial t} \delta_{ij} u_{kk}.
\end{eqnarray}
From Eqs. (\ref{eq:viscosity_lon}), (\ref{eq:viscosity_tra}) and (\ref{eq:viscosity_identity}), we obtain
\begin{eqnarray}
	\lambda' &=& \lambda \gamma, \\
	\mu' &=& \mu \gamma.
\end{eqnarray}
Therefore, we finally obtain the relation
\begin{equation}
	\gamma = \left( \frac{\lambda'}{\lambda} \right)^2 \frac{3 \lambda + 2 \mu}{3 \lambda' + 2 \mu'}.
\end{equation}

\section{\label{sec:app_perturbation-calc} THE DERIVATION OF EQ. (\ref{eq:second_energy})}

Here we derive Eq. (\ref{eq:second_energy}) in Sec. \ref{ssec:discussion_perturbation}.
Because the solution of the unperturbed vibrational mode $\tilde{Q}_{i, n \ell m}^{(0)}$ is given by
\begin{equation}
	\tilde{Q}_{i, n \ell m}^{(0)}(\tilde{t}_\text{vib}) = \frac{1}{\tilde{\omega}_{i, n \ell}}
		\sqrt{2 \tilde{H}_{i, n \ell m}^{(0)}(0)} \sin (\alpha_{i, n \ell m}(0) + \tilde{\omega}_{i, n \ell} \tilde{t}_\text{vib}),
\end{equation}
the vibrational energy coefficient of the second order $\tilde{H}_{i, n \ell m}^{(2)}(\tilde{t}_\text{vib})$ becomes
\begin{eqnarray}
	\tilde{H}_{i, n \ell m}^{(2)}(\tilde{t}_\text{vib})
	&=& \dot{\tilde{Q}}_{i, n \ell m}^{(0)}(\tilde{t}_\text{vib}) \dot{\tilde{Q}}_{i, n \ell m}^{(2)}(\tilde{t}_\text{vib})
		+ \tilde{\omega}_{i, n \ell}^2 \tilde{Q}_{i, n \ell m}^{(0)}(\tilde{t}_\text{vib})
		\tilde{Q}_{i, n \ell m}^{(2)}(\tilde{t}_\text{vib}) \nonumber \\
	&=& \sqrt{2 \tilde{H}_{i, n \ell m}^{(0)}(0)} \{ \dot{\tilde{Q}}_{i, n \ell m}^{(2)}(\tilde{t}_\text{vib})
		\cos (\alpha_{i, n \ell m}(0) + \tilde{\omega}_{i, n \ell} \tilde{t}_\text{vib}) \nonumber \\
	&& + \tilde{\omega}_{i, n \ell} \tilde{Q}_{i, n \ell m}^{(2)}(\tilde{t}_\text{vib})
		\sin (\alpha_{i, n \ell m}(0) + \tilde{\omega}_{i, n \ell} \tilde{t}_\text{vib}) \} \nonumber \\
	&=& - 2 \sqrt{\tilde{H}_{i, n \ell m}^{(0)}(0) \tilde{H}_{i, n \ell m}^{(4)}(\tilde{t}_\text{vib})} \cos (\alpha_{i, n \ell m}(0) + \tilde{\omega}_{i, n \ell} \tilde{t}_\text{vib} - \beta_{i, n \ell m}(\tilde{t}_\text{vib})).
\end{eqnarray}
In the final line, we have used $\tilde{H}_{i, n \ell m}^{(4)} = (\dot{\tilde{Q}}_{i, n \ell m}^{(2)})^2 / 2 + (\tilde{\omega}_{i, n \ell} \tilde{Q}_{i, n \ell m}^{(2)})^2 / 2$, and introduced $\beta_{i, n \ell m}(\tilde{t}_\text{vib})$:
\begin{equation}
	\sin \beta_{i, n \ell m}(\tilde{t}_\text{vib}) = - \frac{\tilde{\omega}_{i, n \ell} \tilde{Q}_{i, n \ell m}^{(2)}(\tilde{t}_\text{vib})}
		{\sqrt{2 \tilde{H}_{i, n \ell m}^{(4)}(\tilde{t}_\text{vib})}}, \qquad
	\cos \beta_{i, n \ell m}(\tilde{t}_\text{vib}) = - \frac{\dot{\tilde{Q}}_{i, n \ell m}^{(2)}(\tilde{t}_\text{vib})}
		{\sqrt{2 \tilde{H}_{i, n \ell m}^{(4)}(\tilde{t}_\text{vib})}}.
\end{equation}

Because the time evolution of the force $\partial \tilde{V}(\tilde{z}_\text{CM}^{(0)}(t), 0) / \partial \tilde{Q}_{i, n \ell m}$ is symmetric around the instant of the collision $t_f / 2$, $\sin \beta_{i, n \ell m}$ at $\tilde{t}_\text{vib} = t_f$ can be written as
\begin{eqnarray}
	\sin \beta_{i, n \ell m}(t_f) &=& \frac{\int_{0}^{t_f} \mathrm{d}t' \frac{\partial \tilde{V}(\tilde{z}_\text{CM}^{(0)}(t'), 0)}
		{\partial \tilde{Q}_{i, n \ell m}} \sin \tilde{\omega}_{i, n \ell} (t_f - t')}{\sqrt{ \left| \int_0^{t_f} \mathrm{d}t' \frac{\partial
		\tilde{V}(\tilde{z}_\text{CM}^{(0)}(t'), 0)}{\partial \tilde{Q}_{i, n \ell m}} e^{i \tilde{\omega}_{i, n \ell} t'} \right|^2}} \nonumber \\
	&=& \frac{\int_{- t_f / 2}^{t_f / 2} \mathrm{d}t' \frac{\partial \tilde{V}(\tilde{z}_\text{CM}^{(0)}(t_f / 2 + t'), 0)}
		{\partial \tilde{Q}_{i, n \ell m}} \sin \tilde{\omega}_{i, n \ell} (t_f / 2 - t')}{\sqrt{\left| \int_{- t_f / 2}^{t_f / 2} \mathrm{d}t'
		\frac{\partial \tilde{V}(\tilde{z}_\text{CM}^{(0)}(t_f / 2 + t'), 0)}{\partial \tilde{Q}_{i, n \ell m}}
		e^{i \tilde{\omega}_{i, n \ell} t'} \right|^2}} \nonumber \\
	&=& \frac{\sin \frac{\tilde{\omega}_{i, n \ell} t_f}{2} \int_{- t_f / 2}^{t_f / 2} \mathrm{d}t' \frac{\partial
		\tilde{V}(\tilde{z}_\text{CM}^{(0)}(t_f / 2 + t'), 0)}{\partial \tilde{Q}_{i, n \ell m}} \cos \tilde{\omega}_{i, n \ell} t'}
		{\left| \int_{- t_f / 2}^{t_f / 2} \mathrm{d}t' \frac{\partial \tilde{V}(\tilde{z}_\text{CM}^{(0)}(t_f / 2 + t'), 0)}
		{\partial \tilde{Q}_{i, n \ell m}} \cos \tilde{\omega}_{i, n \ell} t' \right|} \nonumber \\
	&=& \sin \frac{\tilde{\omega}_{i, n \ell} t_f}{2}.
\end{eqnarray}
In the final expression, we have removed the absolute value in the denominator because $\partial \tilde{V}(\tilde{z}_\text{CM}^{(0)}(t), 0) / \partial \tilde{Q}_{i, n \ell m}$ is always positive and monotonically increase up to $t = t_f / 2$.
We can also calculate $\cos \beta_{i, n \ell m}(t_f)$, and the result becomes
\begin{equation}
	\cos \beta_{i, n \ell m}(t_f) = \cos \frac{\tilde{\omega}_{i, n \ell} t_f}{2}.
\end{equation}
Therefore,
\begin{equation} \label{eq:perturb_beta}
	\beta_{i, n \ell m}(t_f) = \frac{\tilde{\omega}_{i, n \ell} t_f}{2}.
\end{equation}

\newpage %Just because of unusual number of tables stacked at end
\bibliography{dthesis}% Produces the bibliography via BibTeX.

\end{document}